\documentclass[useAMS]{mn2e}
\usepackage{graphicx}

\def\lesssim{\mathrel{\hbox{\rlap{\hbox{\lower4pt\hbox{$\sim$}}}\hbox{$<$}}}}
\def\gtrsim{\mathrel{\hbox{\rlap{\hbox{\lower4pt\hbox{$\sim$}}}\hbox{$>$}}}}

\title[VPHAS+]
{The VST Photometric H$\alpha$ Survey of the Southern Galactic Plane and Bulge (VPHAS+)}
\author[J. E. Drew et al. ]
{J. E. Drew$^1$, E. Gonzalez-Solares$^2$, R. Greimel$^3$,
 M. J. Irwin$^2$, A. Kupcu Yoldas$^2$,
\newauthor
J. Lewis$^2$, G. Barentsen$^1$, J. Eisl\"offel$^4$, H. J. Farnhill$^1$, 
W. E. Martin$^1$, J. R. Walsh$^5$, 
\newauthor
N. A. Walton$^2$, M. Mohr-Smith$^1$, R. Raddi$^6$, S. E. Sale$^7$, 
N. J. Wright$^1$, P. Groot$^8$, 
\newauthor
M. J. Barlow$^9$, R. L. M. Corradi$^{10}$, J. J. Drake$^{11}$, 
J. Fabregat$^{12}$, D. J. Frew$^{13}$, 
\newauthor
B. T. G\"ansicke$^6$, C. Knigge$^{14}$, A. Mampaso$^{10}$, 
R. A. H. Morris$^{15}$, T. Naylor$^{16}$,
\newauthor
Q. A. Parker$^{13}$, S. Phillipps$^{14}$, C. Ruhland$^1$, D. Steeghs$^6$, 
Y.C. Unruh$^{17}$, J. S. Vink$^{18}$, 
\newauthor
R. Wesson$^{19}$, A. A. Zijlstra$^{20}$\\
$^1$School of Physics, Astronomy \& Mathematics, University of Hertfordshire,
College Lane, Hatfield, Hertfordshire, AL10 9AB, U.K.\\
$^2$Institute of Astronomy, Cambridge University, Madingley Road, Cambridge,
CB3 OHA, U.K.\\
$^3$IGAM, Institute of Physics, University of Graz,
Universit\"atsplatz 5, Graz, Austria\\
$^4$Th\"uringer Landessternwarte, Sternwarte 5, 07778, Tautenburg, Germany\\
$^5$ESO Headquarters, Karl-Schwarzschild-Strasse 2, 85748 Garching, Germany\\
$^6$Department of Physics, University of Warwick, Gibbet Hill Road, Coventry, CV4 7AL, U.K>\\ 
$^7$Rudolf Peierls Centre for Theoretical Physics, Keble Road, Oxford, OX1 3NP\\
$^8$Afdeling Sterrenkunde, Radboud Universiteit Nijmegen, Faculteit NWI,
Postbus 9010, 6500 GL Nijmegen, The Netherlands\\
$^{9}$University College London, Department of Physics \& Astronomy, 
Gower Street, London WC1E 6BT, U.K.\\
$^{10}$ Instituto de Astrofisica de Canarias, 38200 La Laguna, Tenerife, Spain\\
$^{11}$Harvard-Smithsonian Center for Astrophysics, 60 Garden Street, 
Cambridge, MA 02138, U.S.A. \\
$^{12}$Observatorio Astr\'onomico, Universidad de Valencia,
Catedr\'atico Jos\'e Beltr\'an 2, 46980 Paterna, Spain\\
$^{13}$Department of Physics \& Astronomy, Macquarie University, NSW 2109, Australia\\
$^{14}$School of Physics \& Astronomy, University of Southampton, Southampton,
SO17 1BJ, U.K.\\ 
$^{15}$School of Physics, Bristol University, Tyndall Avenue, Bristol, BS8 1TL, U.K.\\
$^{16}$School of Physics, University of Exeter, Stocker Road, Exeter, EX4 4QL, U.K.\\
$^{17}$Department of Physics, Blackett Laboratory, Imperial College
London, Prince Consort Road, London, SW7 2AZ, U.K.\\
$^{18}$Armagh Observatory, College Hill, Armagh, Northern Ireland,
BT61 9DG, U.K.\\
$^{19}$European Southern Observatory, Alonso de C\'ordova 3107, Casilla 19001,
Santiago, Chile\\   
$^{20}$Jodrell Bank Centre for Astrophysics, School of Physics \&
Astronomy, University of Manchester, Oxford Road, Manchester M13 9PL, U.K.\\
}

\begin{document}
\maketitle

\begin{abstract}
The VST Photometric H$\alpha$ Survey of the Southern Galactic Plane
and Bulge (VPHAS$+$) is surveying the southern Milky Way in $u, g, r, i$ and
H$\alpha$ at $\sim$1~arcsec angular resolution. Its footprint spans
the Galactic latitude range 
$-5^{\rm o} < b < +5^{\rm o}$ at all longitudes south of the celestial
equator. Extensions around the Galactic Centre to Galactic latitudes 
$\pm 10^{\circ}$ bring in much of the Galactic Bulge.   This ESO 
public survey, begun on 28th December 2011, reaches down 
to $\sim$20th magnitude (10$\sigma$) and will provide single-epoch 
digital optical photometry for $\sim$300 million stars.  The observing 
strategy and data pipelining is described, and an appraisal of the 
segmented narrowband $H\alpha$ filter in use is presented.  Using model
atmospheres and library spectra, we compute main-sequence $(u - g)$, 
$(g - r)$, $(r - i)$ and $(r - H\alpha)$ stellar colours in the Vega
system.  We report on a preliminary validation of
the photometry using test data obtained from two pointings overlapping
the Sloan Digital Sky Survey.  An example of the $(u-g,g-r)$ and
$(r-H\alpha,r-i)$ diagrams for a full VPHAS+ survey field is given.  
Attention is drawn to the opportunities for studies of compact nebulae
and nebular morphologies that arise from the image quality being achieved.
The value of the $u$ band as the means to identify
planetary-nebula central stars is demonstrated by the discovery of the
central star of NGC 2899 in survey data. Thanks to its excellent imaging 
performance, the VST/OmegaCam combination used by this survey is a
perfect vehicle for automated searches for reddened early-type stars,
and will allow the discovery and analysis of compact binaries, white
dwarfs and transient sources.
\end{abstract}


\section{Introduction}
\label{sec:intro}

The $H\alpha$ emission line is well-known as a tracer of 
diffuse ionized nebulae and as a marker of pre- or post-main sequence status
among spatially-unresolved stellar sources.  Since these objects -- both
nebulae and stars -- represent relatively short-lived phases of evolution, 
they amount to a minority population in a mature galaxy like our own.  
Their relative scarcity has in the past stood in the way of developing 
and testing models for these crucial evolutionary stages.  

In the southern hemisphere, the search for planetary nebulae (PNe) has
been served well by H$\alpha$ imaging surveys carried out by the UK
Schmidt Telescope (Parker et al 2005, 2006 and other more recent
works).  Nevertheless, VPHAS+ will have a decisive impact on studies
of complex or smaller nebulae of all types, ranging from
optically-detectable ultra-compact and compact HII regions, to nebulae
around YSOs (including associated jets and HH objects), through 
PNe, to extended emission from D-type symbiotic stars and supernova 
remnants.  The superb spatial resolution, dynamic range, and likely 
photometric accuracy of the VPHAS+ images warrant a step forward in our 
knowledge of the population and detailed characteristics of these object
classes. 

For southern point sources with emission the situation is very
different: there has been little updating of the available catalogues
since the work of Stephenson \& Sanduleak (1971) that was limited to a
depth of 12th magnitude.  The major
groups of emission line stars that remain as challenges to our
understanding include all types of massive star (O stars,
supergiants, luminous 
blue variables, Wolf-Rayet stars, various types of Be star), post-AGB
stars, pre-main-sequence stars at all masses, active stars and compact 
interacting binaries.  Within the disc of the Milky Way, the available 
samples of these objects are typically modest and heterogeneous.
Fixing this deficit via a uniform search of the Galactic Plane for
these rare object classes motivated the 
photometric H$\alpha$ survey of the southern Galactic Plane, first 
proposed for the VLT Survey Telescope (VST) in 2004.  This paper describes 
the realisation of this ESO public survey, now known as the VST Photometric 
H$\alpha$ Survey of the Southern Galactic Plane and Bulge (VPHAS+).

When first proposed, VPHAS (without the plus sign) was envisaged as the 
counterpart to the INT/WFC Photometric H$\alpha$ Survey of the
Northern Galactic
Plane (IPHAS, Drew et al 2005), that had begun in
August 2003.  IPHAS is a digital imaging survey made up of Sloan $r$,
$i$ and narrowband  $H\alpha$ exposures, reaching to $\sim$20th
magnitude, that takes in all Galactic longitudes north of the
celestial equator in the latitude range $-5^{\circ} < b < +5^{\circ}$.
This is all but complete with the new release of a catalogue 
of $\sim$200 million unique objects drawn
from 93 percent of the survey's footprint (Barentsen et al, 2014).  
During the initial VST public survey review process, it was 
agreed that VPHAS should broaden in scope to also incorporate the
Sloan $u$ and $g$ bands (proposed for a separate survey), that are 
particularly useful in picking out OB stars, white dwarfs and other 
blue-excess objects.  With this upgrade to 5 bands, the renamed
VPHAS$+$ became an all-purpose digital optical survey of the southern
Galactic Plane, capable of delivering data at a spatial resolution of
$\sim$1~arcsec or better.  As well as fulfilling the role of 
southern counterpart of IPHAS, VPHAS$+$ is also the counterpart to UVEX, the 
UV-Excess Survey of the Northern Galactic Plane (Groot et al, 2009) that,
at the time of writing, continues on the Isaac Newton Telescope in La Palma.  

The final augmentation of the VPHAS$+$ survey footprint came in 2010
on expanding its footprint to match that of the similarly high 
spatial-resolution near-infrared survey, VISTA Variables in the Via Lactea 
(VVV, Minniti et al 2011).  The survey footprint now includes 
the Galactic Bulge to a latitude of $|b| < 10^{\circ}$, across the
longitude range $-10^{\circ} < \ell < +10^{\circ}$.  The VVV $z$, $Y$,
$J$, $H$ and $K_s$ survey of much of the Bulge and inner Galactic disc
is already complete.  

VPHAS$+$ is poised to become the homogeneous digital
optical imaging survey of the Galactic Plane and Bulge, at $\sim$1~arcsec
angular resolution, that will provide a uniform database of stellar 
spectral energy distributions, from which a range of colour-magnitude 
and colour-colour diagrams of well-established utility can be derived.  
Once calibrated to the expected precision of 2 to 3 percent,
like its northern counterparts, VPHAS$+$ will be quantitatively far
superior to the photographic surveys of the last century, and will
offer significant added value in the form of calibrated narrowband
H$\alpha$ data.  Some of the science enabled is illustrated by the 
studies that the northern surveys, IPHAS and UVEX, have already
stimulated.  These have included a number of works reporting
the discovery of emission line stars, ranging from young objects (e.g.
Valdivieso et al, 2009, Vink et al 2008, Barentsen et al 2011, Raddi
et al 2013) to evolved object classes such as symbiotic and
cataclysmic binaries and compact planetary nebulae (e.g. Corradi et al 
2010, Witham et al 2006, Viironen et al 2009 and Wesson et al 2008).  
The diagnostic power to be expected from the blue $u$ and $g$ bands
has been appraised in a series of papers presenting UVEX and early 
follow-up spectroscopy (Groot et al 2009, Verbeek et al 2012a and 2012b).

\begin{figure*}
\includegraphics[angle=0,width=\linewidth,trim=0 0 0 500]{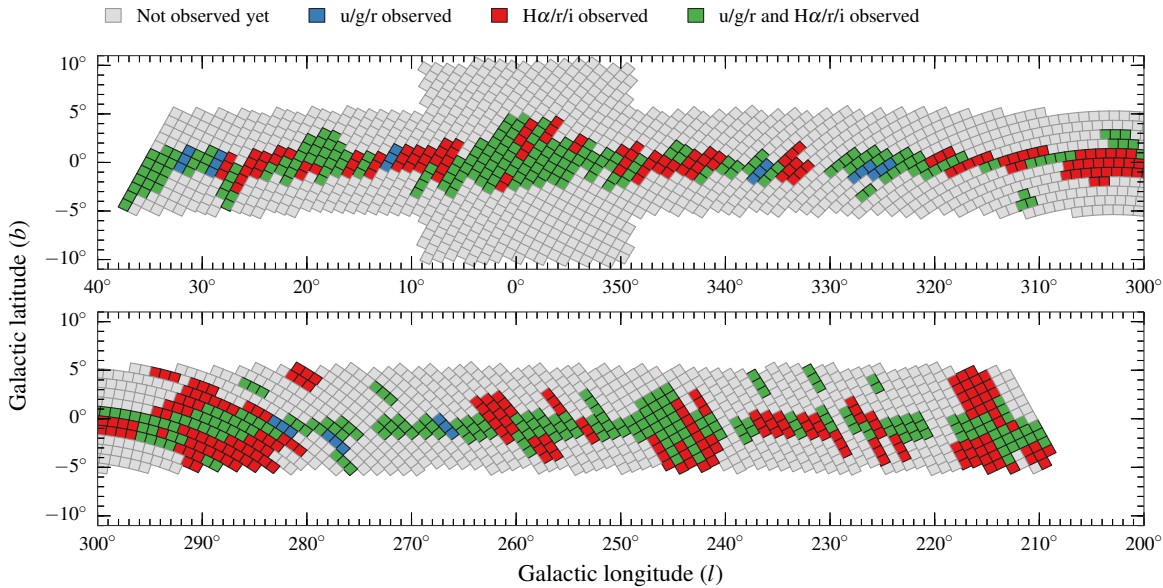}
\caption{The VPHAS$+$ survey footprint plotted in Galactic coordinates.  
  All 2269 fields are shown in outline.  Different colours, as specified in
  the key, are used to identify the observations obtained for each field by 
  1st January 2014 -- essentially 2 years after the start of data-taking.}  
\label{fig:footprint}
\end{figure*}

Any comprehensive survey of the Galactic Plane clearly targets
the main mass component of our own Galaxy, made up of stars, gas and
dust.  When a spatial resolution of 1 arcsec is combined with wide
area coverage spanning 100s of square degrees, as in IPHAS, UVEX and 
VPHAS$+$, it becomes possible to exploit the stellar 
photometry adaptively to solve for the distribution of both the stars
and dust making up the optically-accessible Galactic disk - by means 
that are similar to those already attempted, based on the 
near-infrared 2MASS survey (Drimmel \& Spergel 2001, Marshall et al 2006).   
Methods to achieve 3-dimensional mapping of this kind, now incorporating 
the direct sensitivity of the H$\alpha$ narrowband to stellar 
intrinsic colour, are starting to take shape (Sale et al 2009; Sale 2012).  
This development coincides with the approach of the operations phase of 
Europe's next major astrometric mission, Gaia.  Indeed, the rich 
dynamical picture that Gaia will build of the Milky Way over the next 
decade will be very effectively complemented by stellar energy 
distributions measured for millions of stars from the current
generation of ground-based optical, and near-infrared, wide-field
surveys.  VPHAS$+$, with its haul of photometry in 5 optical bands 
on $\sim$300 million objects, is set to take its place as one of them.
 

This paper presents the main features of VPHAS$+$, including a
description of its execution, the data processing and the nature of the 
photometric colour information it provides.  We begin in the next section 
with a presentation of the observing strategy and the 
data reduction techniques in use.  We then turn to a description and 
evaluation of the narrowband H$\alpha$ filter procured for this survey, in 
section~\ref{sec:ha}.  Following this, in section~\ref{sec:simul},
we present tailored synthetic photometry of main sequence and giant
stars that provides insights into the photometric diagrams that may be 
generated from survey data.  An exercise in photometric validation is 
described in Section~\ref{sec:phot-val} in which Sloan Digital Sky
Survey (SDSS) data are compared with VST observations.  The scene is
then set for an example of VPHAS+ photometry extracted across the
entire square-degree footprint of a single survey field 
(Section~\ref{sec:example}). In Sections~\ref{sec:imaging} and 
\ref{sec:var}, we outline the applications of VPHAS$+$ to spatially-resolved
nebular astrophysics and in the time domain.  The paper ends in 
Section~\ref{sec:last} with a summary, examples of early survey exploitation, 
and a forward look to the first major data release.

\section{Survey observations and data processing}
\label{sec:data}

\subsection{VPHAS$+$ specification}
\label{sec:survey-def}

The footprint of the survey is shown in 
figure~\ref{fig:footprint}.  The OmegaCAM imager (Kuijken 2011) on the VST 
provides a field size of a full square degree, captured on a 4 x 8 CCD 
mosaic.   After allowing for some modest overlap between adjacent fields, we 
arrived at a set of 2269 field centres that will cover the desired Galactic 
latitude band $-5^{\circ} < b < +5^{\circ}$ at all southern-hemisphere Galactic 
longitudes, as well as incorporate the Galactic Bulge extensions to 
$-10^{\circ} < b < +10^{\circ}$ near the Galactic Centre.  The survey footprint 
extends across the celestial equator by a degree or two to achieve an 
overlap with the northern hemisphere surveys IPHAS and UVEX of $\sim$100 
sq.deg. altogether.  This is to create the opportunity for some direct 
photometric cross-calibration.

The target depth of the survey is to reach to at least $\sim$20th magnitude,
at 10$\sigma$, in each of the Sloan $u$, $g$, $r$ and $i$ broadband filters 
and narrowband H$\alpha$.  The bright limit consistent with this goal is
typically 12--13th magnitude.  Presently, all VPHAS+ photometric magnitudes
are expressed in the Vega system.  The original concept was to collect the data 
in all 5 bands contemporaneously, in order to build a uniform library
of snapshot photometric spectral energy distributions for 200 million or 
more stars.  Practical constraints have modified this to the extent that 
the blue filters ($u$, $g$) are observed separately from the reddest ($i$ 
and $H\alpha$), with the $r$ band serving as a linking reference that is 
observed both with $u$,$g$ and $i$,$H\alpha$.  The aim is also to keep the
spatial resolution close to 1 arcsecond.  OmegaCAM and the Paranal site are 
well suited to this in that the camera pixel size is 0.21 arcsec, projected
on sky, and the median seeing achieved is better than 1 arcsec (on occasion 
falling to as little as 0.6 arcsec).

As a means to obtaining better quality control, and to ensure that only a 
minimal fraction of the survey footprint is missed due to the pattern of
gaps between the CCDs in the camera mosaic, every field is imaged at 2 or
3 offset pointings.  This strategy has been carried over from IPHAS
and UVEX in the northern hemisphere, and has the consequence that the 
majority of imaged objects will be detected twice some minutes apart.  In 
the $r$ band, there will be two arbitrarily-separated epochs of data, with 
typically two detections at each of them (i.e. 4 altogether).

\subsection{VPHAS$+$ observations}
\label{sec:obs}

The VST is a service-observing facility, with all programmes queued for 
execution as and when the ambient conditions meet programme requirements.
VPHAS+ survey field acquisition began on 28th December 2011.
Normally the constraint set includes a seeing upper bound of 1.2 arcsec: this 
is only set at a lower, more stringent value for fields expected to present a 
particularly high density of sources (e.g. in the southern Bulge).  In
order that the seeing achieved in the $u$ band is not greatly different from
that in $i$ at the opposite end of the optical range, it is advantageous
to separate acquisition of blue data from red -- hence a split
between 'blue' ($u$,$g$,$r$) and 'red' ($H\alpha$,$r$,$i$) observing 
blocks has been implemented.  This split also permits the use of different
moon distance and phase constraints, such that blue data are obtained when
the moon is less than half full at an angular separation of not less
than 60 degrees, while the limits for red data are set at 0.7 moon
illumination and a minimum angle of 50 degrees.
Avoiding bright-moon conditions is important in order to limit the amount
of moonlight mixed in with diffuse H$\alpha$ emission in the reduced
images. 
No requirement has been placed on the time elapsing between acquisition
of blue and red data.  However, the more forgiving constraints on the 
acquisition of the latter has meant that these are typically executed sooner 
than the former, with the result that many more fields have red
data already than have blue (see fig~\ref{fig:footprint}).  In all 
cases, the final constraint is that the sky is required to be clear, if 
not necessarily fully photometric.

An impressive feature of the camera, OmegaCAM, is its potential to
deliver remarkably undistorted point-source images all the way across 
the 1-degree
field of view. To realise this, it is critical that the VST has an 
actively-controlled primary.   The operational price for this, at the 
present time, is that image analysis and correction has to be carried
out at every filter change or after longer slews.  The overhead added
by this is about 3 minutes.  To reduce the impact of this, 
observations of sets of 3 neighbouring fields are scheduled together,
so that image analysis need only take place every 15-30 minutes -- not 
much more often than would be essential, in any case, to compensate for
the telescope's tracking movement.  As a result, 'contemporaneous' in the 
context of VPHAS+ data-taking means that all 3 blue, or red, filters are 
typically exposed within 40-50 minutes of each other (cf. IPHAS, where 
the more compact camera allows much faster operation, bringing this 
elapsed time down to under 10 minutes).  However, the time difference 
between the blue and red observing blocks for a given field, i.e. the 
$u$/$g$/$r$ data collection and $H\alpha$/$r$/$i$ data collection, can 
be anything from a few hours to more than a year.

\begin{figure}
\includegraphics[angle=0,width=\linewidth,trim= 0 200 0 200]{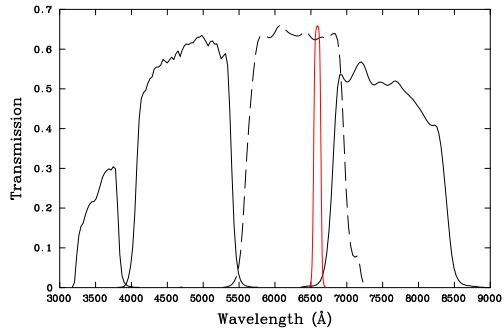}
\caption{The transmission profiles of the Sloan $u$, $g$, $r$ and $i$,
and narrowband $H\alpha$ filters used in all VPHAS$+$ observations. The $r$ 
and $H\alpha$ profiles are shown as a dashed line and in red
respectively just to clearly distinguish them from each other and the
$i$ band. Each profile has been multiplied by the CCD response
function and a model of atmospheric throughput (Patat et al 2011).
The very-nearly grey losses due to the telescope optics (a further
scaling of approximately 0.6) have not been folded in.
}  
\label{fig:filters}
\end{figure}

\begin{figure}
\includegraphics[angle=0,width=\linewidth]{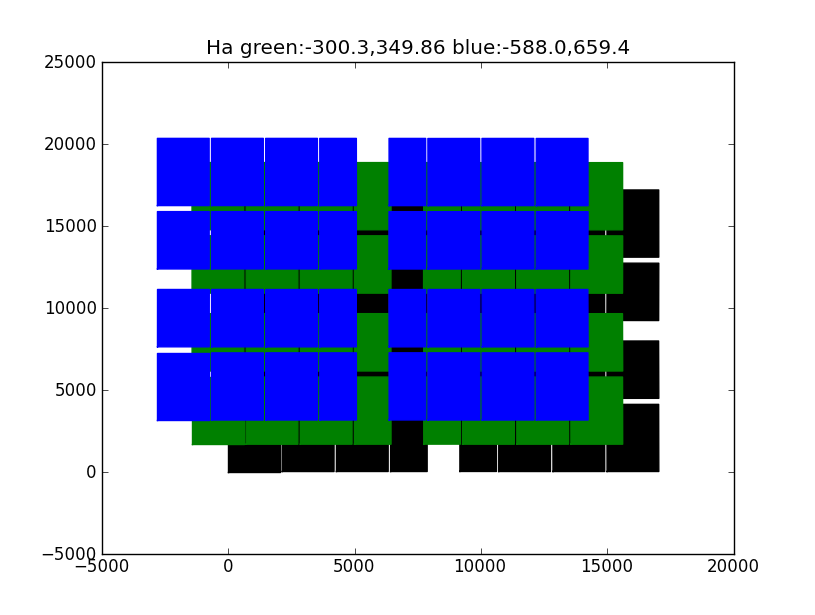}
\caption{An illustration of the VPHAS$+$ offset pattern as it applies
  to the segmented H$\alpha$ filter with extra vignetting due to the T
  bars separating the 4 segments.  The first pointing is to lower
  right, as drawn -- a conservative estimate of the exposed
  unvignetted area is shown in black.  The exposed/unvignetted areas
  of the second and third pointings are shown in green and blue 
  respectively.  The vertical
  and horzontal scales are numbered in pixels (RA {\em increasing} to 
  the right, declination increasing upwards).
}  
\label{fig:offset}
\end{figure}

We provide a reminder of the passbands of the Sloan broadband filters in
use, along with that of the narrowband H$\alpha$, in fig~\ref{fig:filters}.
They are shown scaled by a typical CCD response function and a model of
the atmospheric transmission (Patat et al 2011).  The exposure times
used for the different filters, the number of exposures in each and
the median seeing achieved, up to December 2013, are set out in 
Table~\ref{tab:observations}.  From early on, during the commissioning  
phase of the telescope, it became clear that tracking is
usually good enough that even the 150~sec $u$ exposures, our longest,
do not have to be guided in most circumstances.  Indeed, experience is 
showing that it is safer to rely on the tracking, rather than the
autoguider, to maintain good image quality in the most dense star fields.

\begin{table}
\caption{Observations obtained per survey field.  The median
  seeing quoted is derived from data in hand by December 2013.}
\label{tab:observations}
\begin{minipage}{5.0cm}
\centering
\begin{tabular}{cccc}
\hline
Filter & Exposure     & No. of     & median seeing \\
       & time (secs)  & offsets    & (arcsec)      \\
\hline
\multicolumn{4}{l}{Blue observation blocks}\\
 $u$ & 150 & 2 & 1.01 \\
 $g$ &  40\footnote{Up to 19th February 2013, $g$ exposure times
  were 30 sec.} & 3 & 0.88 \\
 $r$ &  25 & 2 & 0.80 \\
\hline
\multicolumn{4}{l}{Red observation blocks}\\
 $H\alpha$ & 120 & 3 & 0.84 \\
 $r$       &  25 & 2 & 0.82 \\
 $i$       &  25 & 2 & 0.77 \\
\hline
\end{tabular}\par
\vspace{-0.75\skip\footins}
\renewcommand{\footnoterule}{}
\end{minipage}
\end{table}

The pattern of offsets used for each field is illustrated in
fig~\ref{fig:offset}.  The shifts are relatively large, with the outer
pointings differing by $-$588 arcsec in the RA direction and $+$660 arcsec
in declination.  The choices made have largely been driven by the 
characteristics of the narrowband H$\alpha$ filter (discussed below in
section~\ref{sec:ha}), but they also convey the advantage of greatly 
increasing the overlaps between neighbouring fields.  Just the two outermost 
pointings are used when exposing the $u$, $r$ and $i$ filters.  This
leaves 0.4\% of the survey footprint unexposed.  This changes
to complete coverage on including the third intermediate
offset, as is the policy for the $H\alpha$ and $g$ filters.

In accordance with ESO's standard procedures, data are evaluated soon
after collection by Paranal staff and graded before transfer to the 
archive in Garching and to the Cambridge Astronomy Survey Unit
(CASU) in Cambridge.  If the applied constraints are significantly 
violated, the observation block is returned to the queue.

\subsection{Data pipeline}
\label{sec:pipeline}

\subsubsection{Initial Processing}
\label{sec:proc}

\begin{figure*}
\includegraphics[angle=0,width=\linewidth,trim= 0 150 0 0]{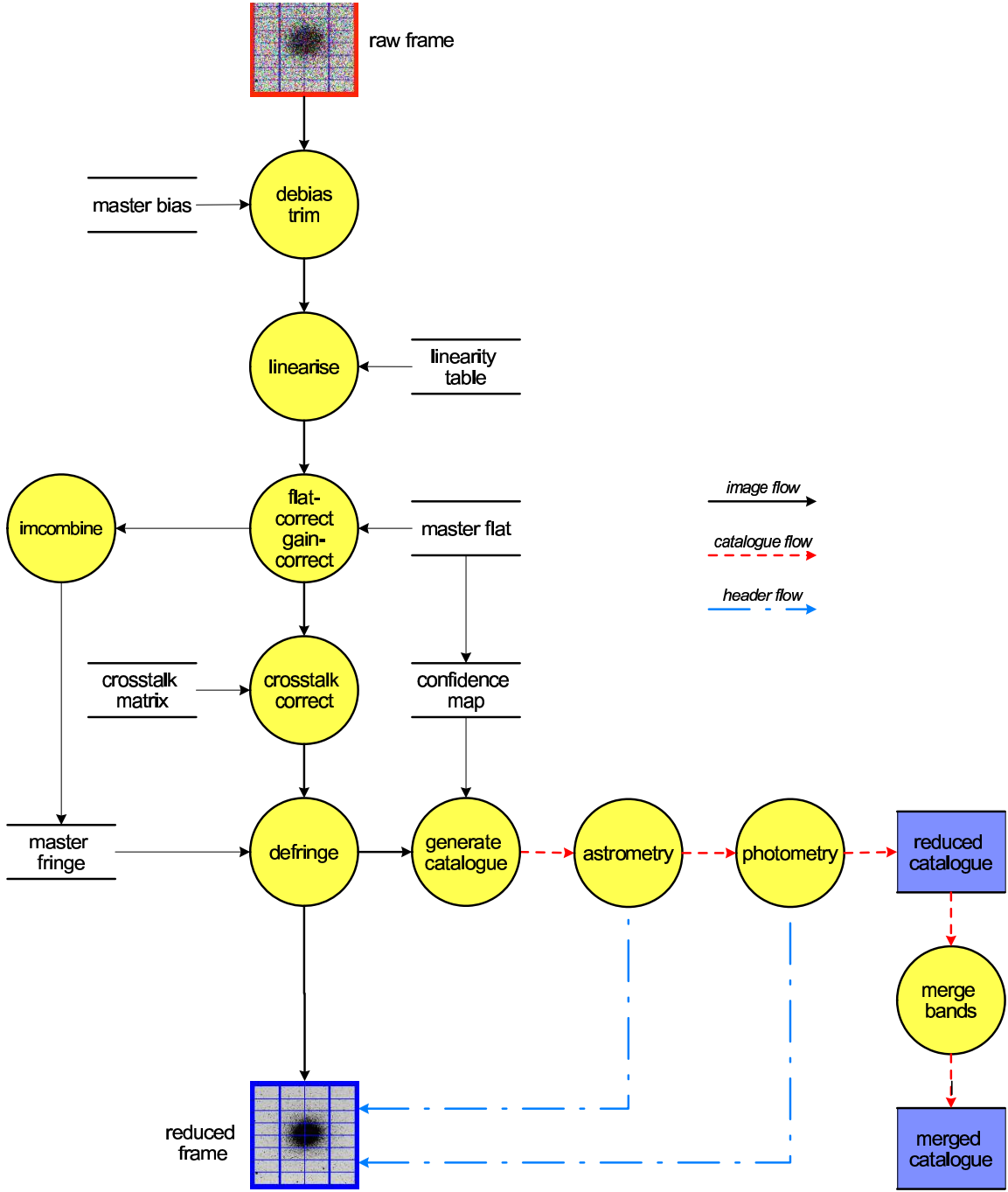}
\caption{Flowchart identifying the main VST data processing steps.  At the
present time all steps up to the reduced single-band catalogue are 
undertaken at CASU.  Band merging is performed at the University
of Hertfordshire.} 
\label{fig:pipeline}
\end{figure*}

From February 29th 2012 raw VST data have been routinely transferred from 
Paranal to Garching over the Internet.  For each observation the imaging data 
are stored in a Multi-Extension FITS file (MEF) with a primary header 
describing the overall characteristics of the observation (pointing, filter, 
exposure time, etc.) and thirty two image extensions, corresponding to
each of the CCD detectors, with further detector-level information in
the secondary
headers. The 32-bit integer raw data files are Rice-compressed at source 
using lossless compression (e.g. Sabbey 1998).  
The files are then checked and ingested into 
the ESO raw data archive in Garching. As soon as the data for any given night 
become available they are automatically transferred to the Cambridge 
Astronomy Survey Unit (CASU) for further checks and subsequent
processing. The VST web pages at CASU provide an external interface
for both monitoring processing status 
(http://casu.ast.cam.ac.uk/vst/data-processing/) and overall survey 
progress and access (http://casu.ast.cam.ac.uk/vstsp/).

The processing sequence is similar to that used for the IPHAS survey
of the northern Galactic Plane (e.g. Gonzalez-Solares et al 2008), 
while the higher level control software is based on that developed for the 
VISTA Data Flow System (VDFS, Irwin et al 2004). Here we briefly outline the 
processing steps illustrated in figure~\ref{fig:pipeline}, emphasising
the main differences relative to the current VDFS standard.  A more detailed 
description of the VST processing pipeline is currently in preparation 
(Yoldas et al 2014).

Science images are first debiassed. Full two-dimensional bias removal is 
necessary due to amplifier glow during readout being present in some 
detectors.  The master bias frames are updated daily from calibration
files taken as part of the operational cycle.  The OmegaCAM detectors
are linear to better than 1\% over their usable dynamic range removing
the need for a linearity correction.  Hence this stage in the pipeline 
processing (figure~\ref{fig:pipeline}), although part of the pipeline 
architecture, is currently bypassed.  

Flatfield images in each band
are constructed by combining a series of twilight sky flats obtained
in bright sky conditions.  The timescale to obtain sequences of these
for all deployed filters is typically one to two weeks.  So as to 
adequately trace the variations in the pattern and level of scattered 
light in these flats, the master flats derived from them are updated
on a monthly cycle (how these are corrected for scattered light
is described in Section~\ref{sec:illcorr}).
Four of the detectors, in extensions 29--32, suffer from
inter-detector cross-talk, whereby saturated bright stars in one detector
can cause noticeable positive or negative low level ($\approx$0.1\%) ghost 
objects in adjacent detectors.  The impact of these is minimised in the 
pipeline by applying a pre-tabulated cross-talk correction matrix to each 
of the affected images.

The flatfield sequences plus bad pixel masks are used to generate the 
confidence (weight) maps (e.g. Irwin et al 2004)
used later during catalogue generation and any subsequent image stacking or 
large area mosaicing. After flatfielding, science images generally
have well behaved sky backgrounds which makes subsequent image
processing straightforward.  Where direct scattered light is present
in them, it is an additive phenomenon that is dealt with automatically 
during object catalogue generation. 

\begin{figure*}
{\includegraphics[angle=-90,width=0.45\textwidth]{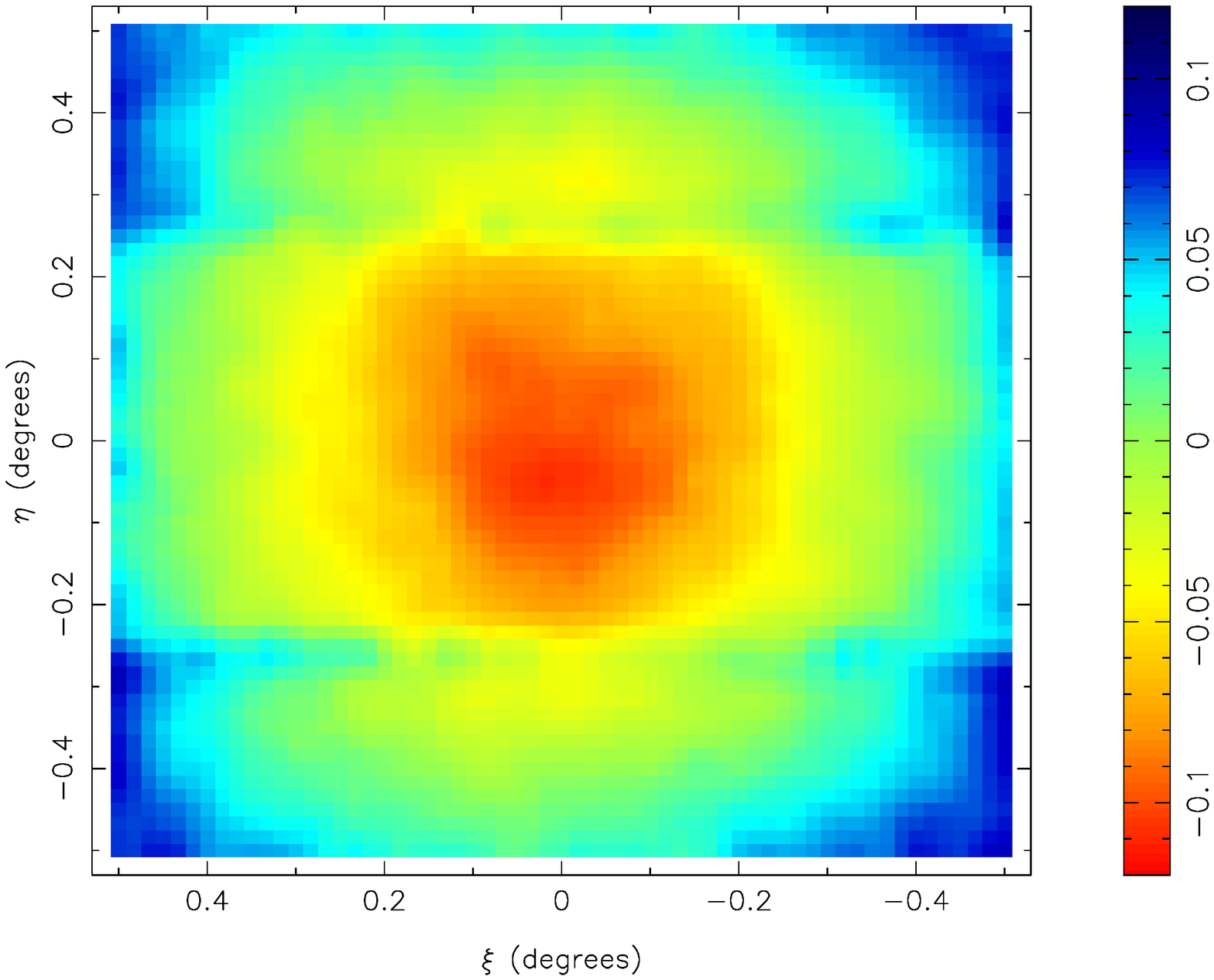}}
{\includegraphics[angle=-90,width=0.45\textwidth]{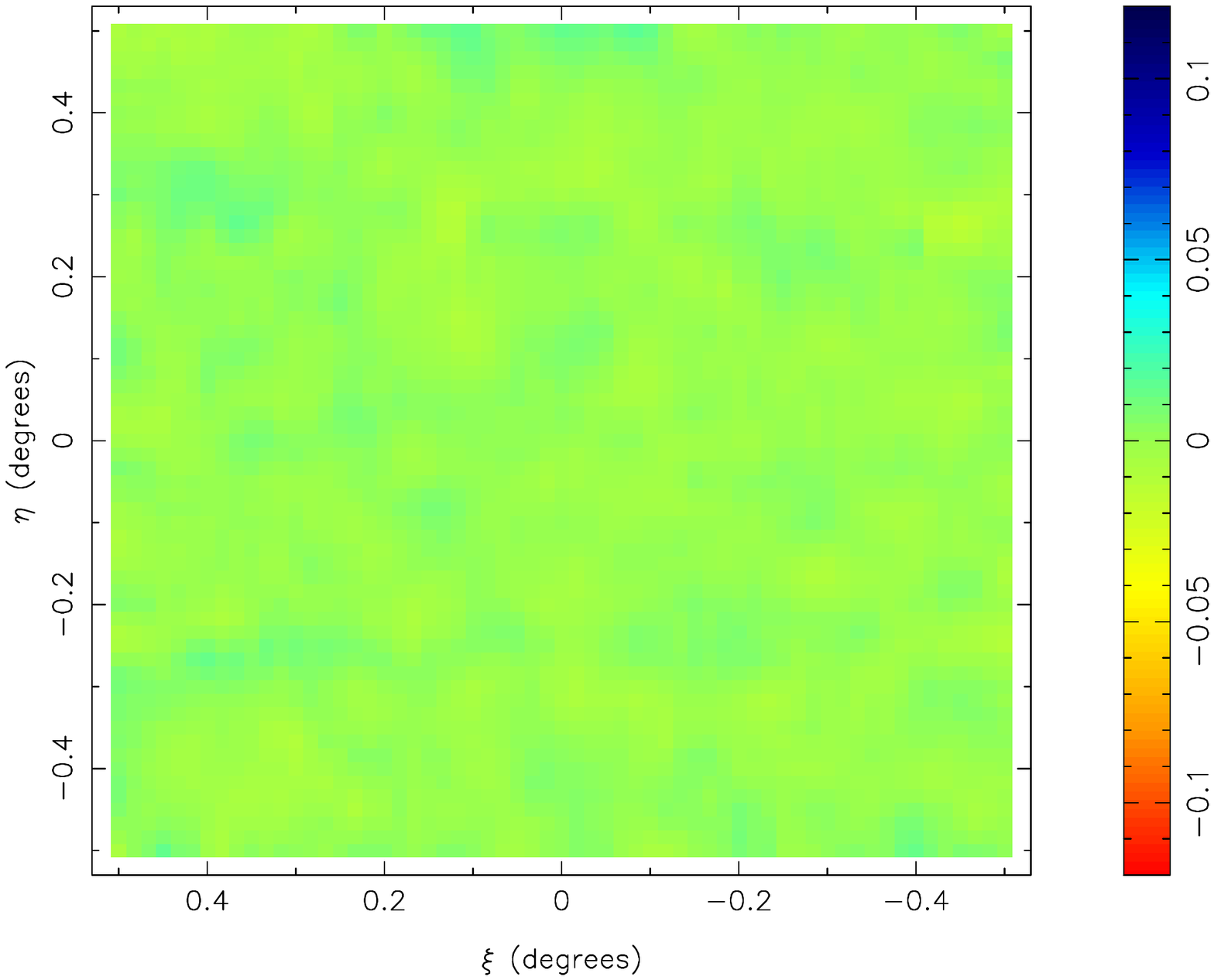}}
\caption{The left-hand panel shows an example of the deduced scattered
light component present in June 2013 $r$-band data.  The right-hand panel
shows the corrected outcome.  These maps were constructed using 286318
APASS object matches over the square-degree field.  Before correction
scattered light gradients amounting to a 20 to 25 percent variation from
corners to centre are present.  This flattens to around $\pm$2 percent.
} 
\label{fig:ill-corr}
\end{figure*}

The redder passbands used in VST observations, in this case the $i$-band,
show fringing patterns at $\approx$2\% of the sky background level.
Defringing is done using a standard CASU procedure (Irwin \& Lewis 2001). 
The fringe frames used are derived from other VST public 
survey data taken as close as possible in time at higher Galactic latitude.  
This approach works because the fringe pattern induced by sky emission lines
at Paranal is quite stable over long periods.  The fringe frames
are automatically scaled and subtracted from each science image reducing
the residual fringing level to well below the sky noise.

Catalogue generation is based on IMCORE\footnote{Software publicly
available from  http://casu.ast.cam.ac.uk} (Irwin, 1985) and makes 
direct use of the confidence maps, derived from the flat fields, to 
suitably weight down unreliable parts of the images.  This step includes 
object detection, parameterization
and morphological classification, together with generation of a range
of quality control information.  Because of the extensive presence of 
diffuse emission throughout the southern Galactic Plane, particularly
in H$\alpha$, a version of each affected image is cleaned of
nebulosity using the NEBULISER\footnote{Software publicly available 
from http://casu.ast.cam.ac.uk, see also Irwin (2010)} 
 for the purpose of catalogue 
generation only.  This achieves a more careful removal of background
and ultimately leads to more complete and, on average, more faithful
object detection than in the absence of this step.  

With object catalogues available for every VPHAS+ survey image, it is then 
possible to improve the rough World Coordinate System (WCS) based on 
the telescope pointing and general system characteristics.  The WCS is
progressively refined using matches between detected objects and the 2MASS 
catalogue (Skrutskie et al 2006).  Despite the large field of view, the VST
focal plane is almost free of distortion, and a standard tangent plane
projection yields residual systematics of $\sim$25mas over the entire field.

\subsubsection{Photometric Calibration}
\label{sec:photcal}

Provisional photometric calibration is based on a series of standard star 
fields observed each night (e.g. Landolt 1992).  For each night a zeropoint 
and error estimate using the observations of all the standard fields in each 
filter is derived.  The flatfielding stage nominally places all detectors on 
a common internal gain system implying, in principle, that a single zeropoint 
suffices to characterise the whole focal surface.  Colour equations
are used to transform between the passbands in use on the
VST and the Johnson-Cousins system of the published standard-star 
photometry.  The calibration is currently in a VST system that uses the
SED of Vega as the zero-colour, almost zero-magnitude, reference object. 

The $u$ band data are the most challenging to calibrate.  As this
part of Vega's spectrum, and also the average standard star plus the detector 
reponse, are falling rapidly it would be surprising if there were no 
offsets in $u$ due to nonlinearities in the required colour transforms and,
perhaps, to degenerate colour transforms for hotter stars. Early
experience of working with $u$ data do indeed suggest that offsets of 
up to a few tenths of a magnitude are sometimes present (see 
Sections~\ref{sec:phot-val} and \ref{sec:example}).

The colour transforms currently in use to define the VPHAS internal
system are given below.  

\begin{eqnarray*}
	u_{VST} & = & U + 0.035 \, (U - B) \\
	g_{VST} & = & B - 0.405 \, (B - V) \\
	r_{VST} & = & R + 0.255 \, (V - R) \\
	i_{VST} & = & I + 0.215 \, (R - I) \\        
  H\alpha_{\,VST} & = & R + 0.025 \, (V - R) \\
\end{eqnarray*}

The transform for the narrowband H$\alpha$ is an approximate {\em
  initial} solution needed for the subsequent illumination correction
stage.  At catalogue bandmerging this is superceded (see 
Section~\ref{sec:simul}). 

\begin{figure*}
\resizebox{8cm}{8cm}{\includegraphics{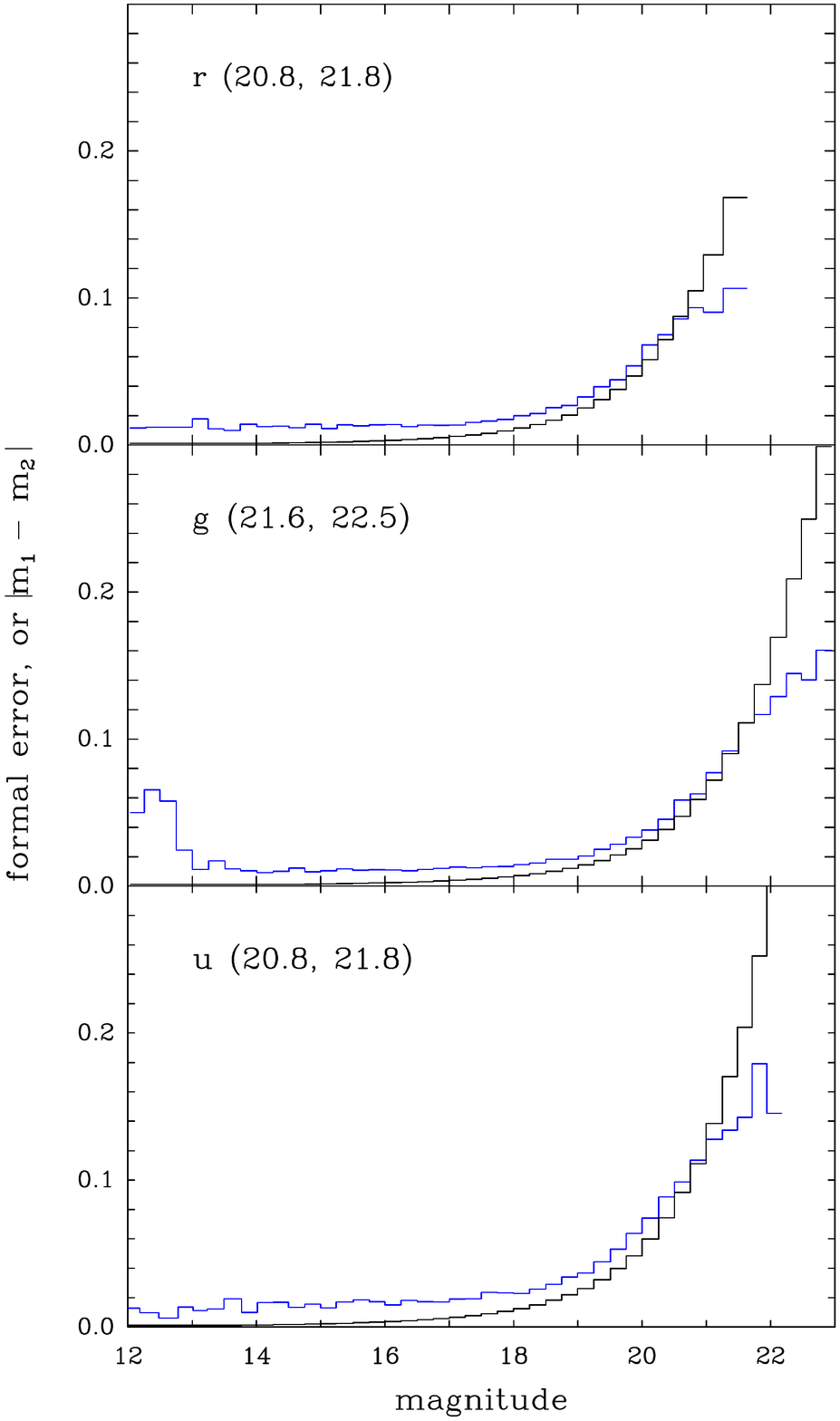}}
\resizebox{8cm}{8cm}{\includegraphics{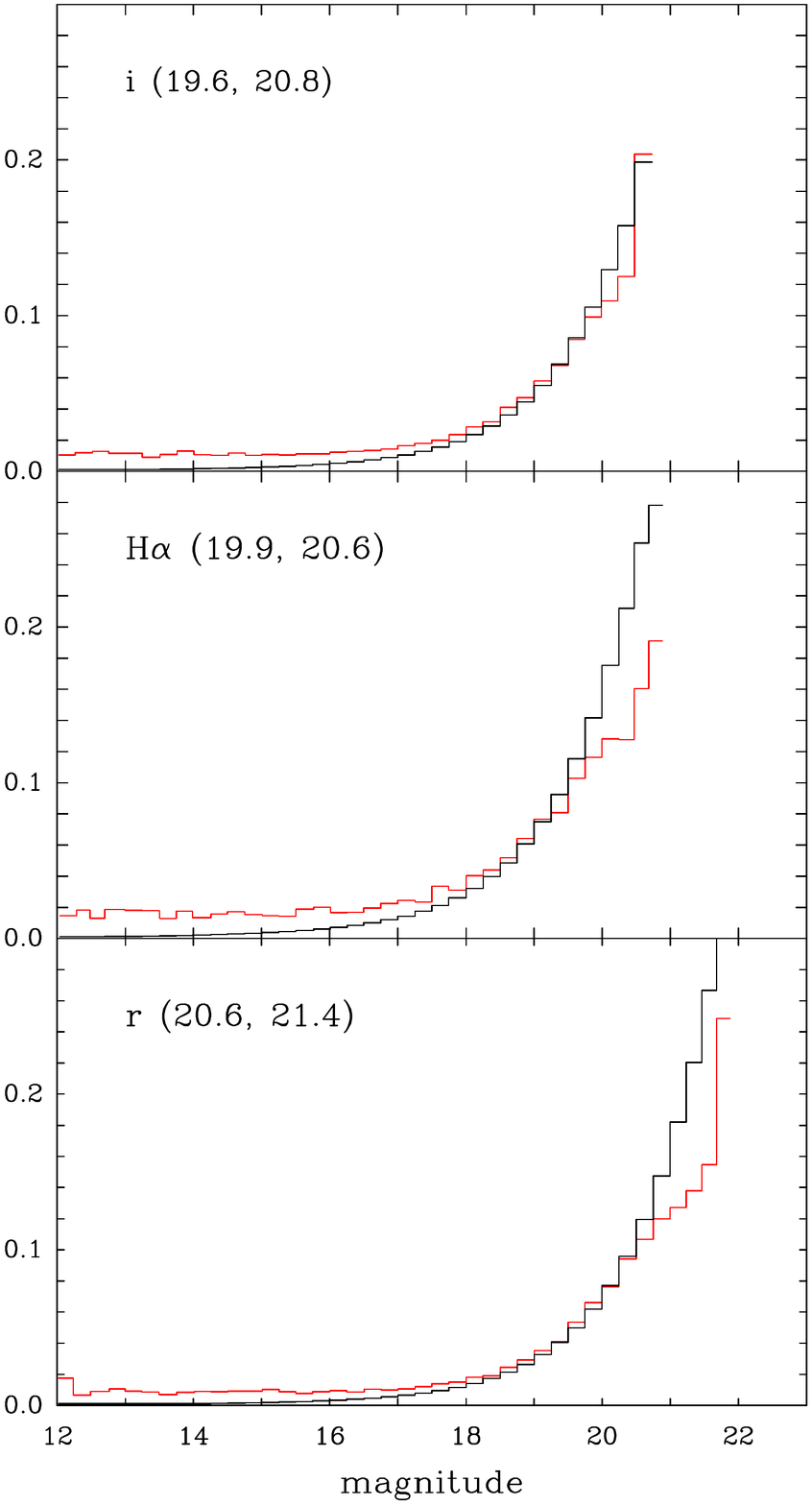}}
\caption{Photometric errors in VPHAS$+$ data as a function of 
magnitude.  All data are drawn from an area of $\sim$0.2~sq.deg in 
field 1679, positioned $\sim$20 arcmin E of Westerlund 2.  The left hand 
panel refers to blue observing block (OB) data, while the right
refers to red OB data.  The coloured
histogram in each component plot shows the mean absolute deviation of
the magnitude difference, $(m_1 - m_2)$, per 0.25-magnitude bin, while 
the bin means of the suitably-corrected pipeline estimate of the
random error on this difference are in black.  The 10- and 5-$\sigma$ 
magnitude limits are specified in the brackets next to each filter
name.  The rising observed mean deviation seen in $g$ at the bright
end is due to the onset of saturation.} 
\label{fig:errors}
\end{figure*}

\subsubsection{Illumination Correction}
\label{sec:illcorr}

The main difficulty in deriving an accurate photometric calibration
over the one degree field arises from the multiplicative systematics 
caused by scattered light in the flatfields.  The VST (at least up to the 
introduction of baffles early in 2014) has proved particularly
susceptible to variable scattered light.  Its impact has varied from 
month to month depending on conditions prevalent at the time the 
flat-field sequences were taken.  An illustration of the amount and 
character of the master-flat correction required is provided in 
figure~\ref{fig:ill-corr}.

The scattered light is made up of multiple components with different 
symmetries and scales.  These range from $\approx$10 arcsec with x-y
rectangular symmetry, e.g. due to scattering off masking strips above the
CCD readout edges, to large fractions of the field due to radial concentration
of light in the optics and to non-astronomical scattered light entering
obliquely in flatfield frames. After some experimentation, and external
verification, we found that the APASS all-sky photometric g,r,i catalogues
(http://www.aavso.org/apass) provide a reliable working solution 
to the illumination correction problem inherent in VST data 
(see fig~\ref{fig:ill-corr}).  These catalogues also provide an independent 
overall photometric calibration tied to the SDSS AB magnitude system and
will be used in future updates to define an alternative finer-grained
temporal AB magnitude zeropoint.

All filters used are treated in the same manner with colour equations 
set up to define transformations between the APASS g,r,i SDSS-like
calibration and the VPHAS+ u,g,r,i,H$_{\alpha}$ internal system.

Illumination corrections are re-derived for each filter once a month.
Application of these corrections via the master flats reduces the 
residual systematics across the entire field to below the 1\% level
for the broadband filters and to within 2\% for the segmented
narrowband H$\alpha$, except in vignetted regions (see Section~\ref{sec:ha}).

\subsubsection{Quality Control}
\label{sec:qc}

In addition to the usual VDFS quality-control monitoring of average stellar 
seeing and ellipticity, sky surface brightness and noise properties,
we have also initiated a more detailed analysis of the image
properties based on inter-detector comparisons.

The well-aligned coplanar detector array coupled with the curved focal surface
is extremely sensitive to imperfections in focus which are relatively easy to
detect using the detector-level average seeing measurement variation 
available for each of the 32 detectors.  Likewise the variation in
average stellar ellipticity from each detector over the field is used
to monitor rotator angle tracking problems.

All of this information can be used in addition to the observation
block (OB) grades provided by ESO and is incorporated within all data 
product files and also the progress database.

\subsection{Limiting magnitudes and errors}
\label{sec:errors}

The present convention for VPHAS+ and this paper is that all
magnitudes are expressed in the Vega system, which imposes zero
intrinisc colour for A0 stars.  The 5-sigma limiting magnitudes
commonly achieved per exposure range from 20.5--21.0 for H$\alpha$ up to 
22.2--22.7 for Sloan $g$.  The 10$\sigma$ limits are about 1 magnitude 
brighter. 

Every source flux or magnitude determined via the pipeline has a
formal error associated with it.  We provide an example of how these 
compare with empirical magnitude differences, by extracting a sample
of stars from a 0.25 sq.deg catalogue, cut out from the survey field
including Westerlund 2 (field 1679, see also sec~\ref{sec:example}) in
order to examine the pattern of errors (fig~\ref{fig:errors}). 
The sky area chosen is offset from the cluster to the east by $\sim$20 
arcmin and exhibits moderate diffuse ionised nebulosity.  In the
southern Plane, the presence of some nebulosity, particularly 
affecting $r$ and $H\alpha$ exposures, is more the rule than the exception.  
Sources classified as probable stars in both $g$ and $r$ (blue filter set, 
left panel) and $r$ and $i$ (red filter set, right hand panel) in two 
consecutive offset exposures have been selected.  The selection also 
required that each extracted magnitude was unaffected by vignetting
and bad pixels (confidence level $> 95$).  This step is particularly 
important for H$\alpha$ given the extra vignetting introduced by the 
cross bars of the segmented filter (see Section~\ref{sec:ha}).  

The faintest stars that might have been included in the plots for $i$
and blue $r$ (top row in fig~\ref{fig:errors}) are absent because of a 
requirement that every included source should {\em also} be picked up
in, respectively, red $r$ and $g$.  Fainter objects than the apparent 
limits certainly exist in these bands.  This feature follows directly
from the typically red colours 
of Galactic Plane stars at magnitudes fainter than $\sim$13 that are
the target of this survey.  For the same reason, it is not uncommon
for the $u$ band source counts to be one or more orders of magnitude lower
than those of the $i$ band.  The role of the $u$ band is to pick out
the unusual rather than to characterise the routine.

To bring out the systematic effects present, the specific
comparison made in fig~\ref{fig:errors} is between the bin means of
the absolute magnitude differences, $|m_1 - m_2 - \delta|$, between the two 
exposures, and the expected random error on the difference
derived from the pipeline rms errors on the individual magnitude 
measurements.  The quantity, $\delta$, is the median magnitude
difference computed from all bright stars down to 18th magnitude 
($r$, $i$ and H$\alpha$), or 19th magnitude ($u$ and $g$).  This was
small in all cases -- the largest value being 0.011 for $u$.  On the
other hand, the correction applied to the pipeline errors was, first, 
to multiply the single-measurement magnitude error by $\sqrt{2}$ to
give the rms error on the $(m_1 - m_2)$ difference, and then to
multiply by $\sqrt{2/\pi}$ in order to convert the measure of
dispersion from rms to a mean deviation.    

At magnitudes brighter than 18--19 in fig~\ref{fig:errors}, the
scatter in the empirical results can be seen to be appreciably greater 
than that 'predicted' for the random component by the pipeline.  The 
scale of the difference indicates that a further error component of 
$\sim$0.01--0.02 magnitudes is present.  The amount and
filter-dependence of these levels of error are entirely consistent 
with the uncertainties estimated above for the flatfield and illumination 
corrections: as noted in section~\ref{sec:illcorr}, the VST is 
presently prone to quite high and variable levels of scattered light.  
Practical remedies for this are under consideration by ESO -- when
implemented these should tighten up the error budget.  

The enhanced mean magnitude difference seen for $g < 13$ in 
fig~\ref{fig:errors} is typical of what is seen as saturation effects 
begin to set in.  For all the other bands, in this example, saturation 
sets in at magnitudes a little brighter than 12th.  A safe working 
assumption across VPHAS+ would be that saturation is never troublesome
at magnitudes fainter than 13, but always an issue for magnitudes 
brighter than 12.  

Figure~\ref{fig:errors} identifies the 5$\sigma$ and 10$\sigma$ magnitude
limits for each filter achieved in this representative example.  The
seeing at the time of these observations, as measured in the pipeline, 
ranged from 0.8 to 1 arcsec (cf Table~\ref{tab:observations}).

\section{The narrowband $H\alpha$ filter}
\label{sec:ha}

\subsection{Overview}

\begin{table*}
\caption{Summary of filter segment properties}
\begin{tabular}{lccccr}
\hline
Segment & sky quadrant & CCDs covered & centre, mean, corner CWL & mean integrated \\
        &   &   & (\AA ) & throughput (\AA ) \\
\hline
A & SW &  1 -- 8  & 6580.2 -- 6585.4 -- 6595.3 & 98.64 \\  
B & SE & 17 -- 24 & 6596.1 -- 6585.4 -- 6578.9 & 103.38 \\ 
C & NE & 25 -- 32 & 6582.8 -- 6591.9 -- 6599.8 & 99.74 \\ 
D & NW &  9 -- 16 & 6581.7 -- 6594.3 -- 6603.8 & 99.49 \\ 
\hline
\end{tabular}
\label{tab:segments}
\end{table*}

\begin{figure}
\centering{\includegraphics[angle=0,width=0.8\linewidth]{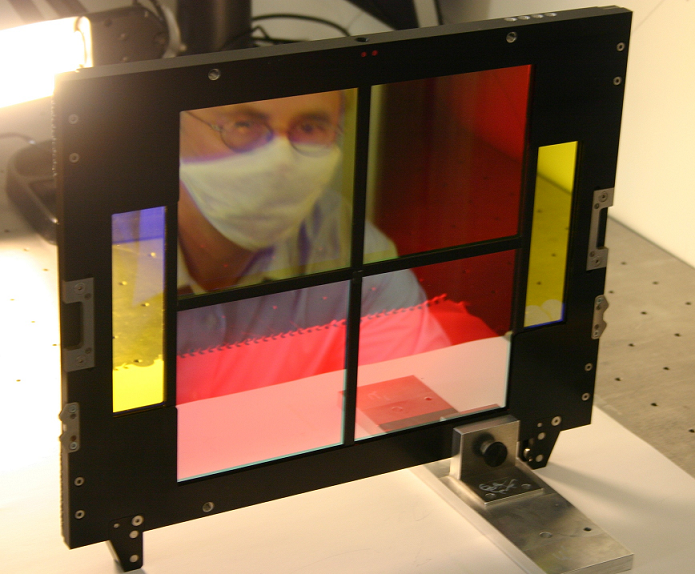}}
\caption{The segmented H$\alpha$ filter, photographed in the lab soon
after receipt and just prior to measuring its transmission.  The filters
to either side of the 2$\times$2 array of H$\alpha$ segments, transmit 
as $r$-band and cover the guide CCDs.}  
\label{fig:hafilter}
\end{figure}

A filter required to select a narrow band across a large 27$\times$27
cm$^2$ image plane is a challenging fabrication problem.  At the 
telescope, the filter in use for VPHAS$+$ is known as NB-659.
At the time it was commissioned in 2006, the purchase of a single-piece 
narrowband filter was offered only by one supplier and was well beyond 
budget. This left the 4-segment option as the achievable alternative.

\begin{figure}
\centering{\includegraphics[angle=0,scale=0.5,trim= 0 240 0 150]{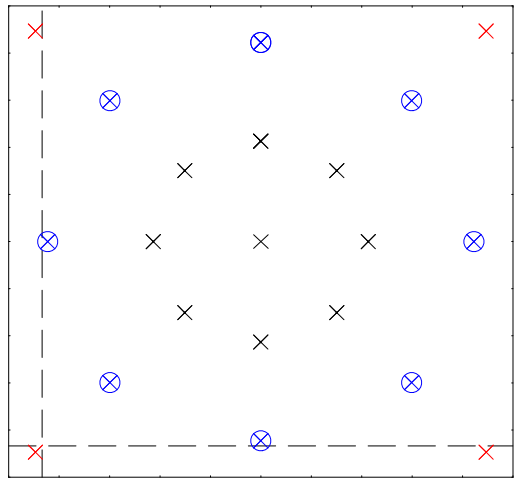}}
\caption{A map of the positions within each filter segment at which
  the transmission was measured. The colour codings are used again in
  later figures to distinguish the central positions (black, 
crosses, sampling roughly 30\% of the segment area), intermediate radii
(blue, encircled crosses, sampling $\sim$half the area), and corners 
(red crosses, 15\% of the area -- of which almost half is lost to 
vignetting).  The dashed lines define the limits of the strips
2~arcmin wide that experience any vignetting due to the filter T-bars.  
They are drawn here as for segment D.}  
\label{fig:segment-map}
\end{figure}

The H$\alpha$ filter was constructed based on a specification supplied 
by the OmegaCAM consortium, setting as goal a central wavelength of 6588 \AA ,
and a bandpass of 107 \AA .  It was delivered in the summer of 2009,
and was shortly thereafter tested at the University of Munich
Observatory, using the optical lab set up by the OmegaCAM consortium for 
filter testing.  A photo of the filter at that time is shown in 
figure~\ref{fig:hafilter}.  The transmission of each filter segment
was measured at 21 positions forming 
a coarse radial pattern (fig~\ref{fig:segment-map}) using a 
monochromator beam adjusted to emulate the f-ratio 5.5 VST/OmegaCAM optical 
system.  The logic of the chosen measurement pattern is to give a good 
sampling of the dominantly radial variation ofthe transmission profile
due to the turntable rotation in the filter coating chamber.  The
diameter of the monochromator beam used in the measurements
was 4-5 mm.  This is a more compact beam than that of starlight at the 
telescope, which fills a spotsize of up to 12 mm on passing through the 
filter out of focus. Consequently, the actual performance will be a somewhat 
areally-smoothed version of the performance revealed by the lab measurements 
and their subsequent simulation.   
The filter was shipped to Paranal and VST in the spring of 2011, after 
some final selective remeasuring.  These confirmed there had been no 
discernible bandpass changes in store since delivery almost 2 years earlier.  

\begin{figure}
\includegraphics[angle=0,scale=0.4,trim= 0 180 0 180]{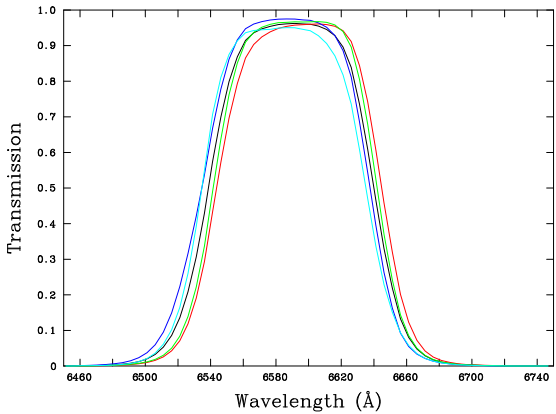}
\caption{The mean transmission profiles of the individual glass 
segments, A to D (cyan, blue, green and red respectively), making up the 
$H\alpha$ filter and the overall mean profile (black).}  
\label{fig:transmission}
\end{figure}

At the time the monochromator measurements were made, a segment naming 
scheme was put in place (segments A, B, C and D) which is re-used here.  
Presently the filter is housed in magazine B of OmegaCAM, which means that 
in terms of the view of the sky, segment A spans the SW section of the image 
plane, B the SE, while C and D span the NE and NW respectively. 
Table~\ref{tab:segments} 
identifies the mosaic CCDs beneath each segment, and sets down the 
centre-to-corner range in central wavelength (CWL) and the typical 
throughput integral.  The laboratory tests showed us that the CWL of 
segments A, C and D is shortest in the segment centre, and drifts
longwards according to a centro-symmetric pattern, 
as the corners and sides are approached. 
For segment B, the centre-to-corner drift is reversed, with
the result that the corner CWLs are bluer than in the centre of the
glass.  Segment B also has the highest mean FWHM, and highest average peak 
transmission: integrated over the bandpass this is a difference in throughput 
of 0.045 magnitudes relative to A, C and D.  The pipeline-applied 
illumination correction aims to eliminate this contrast.
Area-weighted transmission profiles for the 4 segments are shown in 
fig~\ref{fig:transmission}, along with the overall mean profile.  The 
latter is also given numerically in table~\ref{tab:filter-mean}.

\begin{table}
\caption{Mean transmission for NB-659}
\begin{tabular}{cccc}
\hline
Wavelength & Transmission & Wavelength & Transmission\\
  (\AA )   &              &  (\AA )    &             \\
\hline
  6456.3 & 0.000 & 6591.5 & 0.962 \\ 
  6461.3 & 0.001 & 6696.5 & 0.961 \\ 
  6465.8 & 0.001 & 6601.0 & 0.960 \\ 
  6470.8 & 0.002 & 6616.0 & 0.955 \\ 
  6475.9 & 0.002 & 6611.0 & 0.945 \\ 
  6480.9 & 0.003 & 6616.1 & 0.928 \\ 
  6485.9 & 0.005 & 6621.1 & 0.896 \\ 
  6490.9 & 0.008 & 6626.1 & 0.839 \\ 
  6496.0 & 0.012 & 6631.1 & 0.736 \\ 
  6501.0 & 0.020 & 6636.2 & 0.609 \\ 
  6506.0 & 0.033 & 6641.2 & 0.466 \\ 
  6511.1 & 0.053 & 6646.2 & 0.230 \\ 
  6516.1 & 0.086 & 6651.2 & 0.219 \\ 
  6521.1 & 0.136 & 6656.3 & 0.133 \\ 
  6526.1 & 0.208 & 6661.3 & 0.081 \\ 
  6531.2 & 0.307 & 6666.3 & 0.048 \\ 
  6536.2 & 0.429 & 6671.3 & 0.029 \\ 
  6541.2 & 0.575 & 6676.4 & 0.018 \\ 
  6546.3 & 0.700 & 6681.4 & 0.011\\ 
  6551.3 & 0.799 & 6686.4 & 0.007\\ 
  6556.3 & 0.868 & 6691.5 & 0.005\\ 
  6561.4 & 0.915 & 6696.5 & 0.003\\ 
  6566.4 & 0.936 & 6701.5 & 0.002\\ 
  6571.4 & 0.947 & 6706.5 & 0.002\\ 
  6576.4 & 0.954 & 6711.5 & 0.001\\ 
  6581.5 & 0.959 & 6716.5 & 0.001\\ 
  6586.5 & 0.961 & 6721.6 & 0.000\\ 
\hline
\end{tabular}
\label{tab:filter-mean}
\end{table}

Compared to the H$\alpha$ filter used in the IPHAS survey, NB-659 has
a CWL that is redder on average by $\sim 20$~\AA , it is around 10
percent wider, and has a higher overall throughput leading to
zeropoints $\sim$0.2 higher.  The known variations of bandpass across
the 4 segments has implications for how best to exploit VPHAS+
data.  To anticipate these we have carried out two types of simulation
based on the lab measurements in order to identify them. We describe 
these next, and summarise the implications in Section~\ref{sec:ha-final}.

\subsection{Simulation of the main stellar locus in the $(r - H\alpha, 
r - i)$ diagram}
\label{sec:MSHa-sim}

\begin{figure}
\centering{\includegraphics[angle=0,scale=0.45,trim= 0 350 0 0]{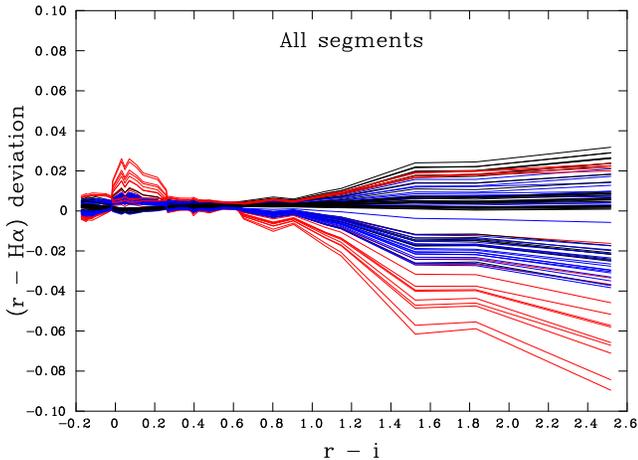}}
\caption{The r - H$\alpha$ deviations computed for all measured
positions on the H$\alpha$ filter, NB-659, after correcting the data for 
bandpass integrated throughput variations.  The data are colour-coded
according to position of measurement as in fig~\ref{fig:segment-map}.
The most discrepant corner positions, all plotted in red, are located
in segments C and D.}  
\label{fig:track-comparison}
\end{figure}

To gain an impression of the extent of the uniformity of performance
with regard to normal main sequence stars, ($r - H\alpha, r- i$) tracks 
were (1) computed for each measured $H\alpha$ transmission profile using 
exactly the same method as was followed by Drew et al (2005) for the analysis 
of IPHAS data, (2) rescaled to a common integrated throughput, mimicking the 
effect of the pipeline illumination correction, (3) compared to the mean 
pattern by subtracting off the computed mean track.  The result of this is 
shown as fig~\ref{fig:track-comparison}.  The track differences
picked out in red are from the segment corners exhibiting the largest CWL 
shifts.  It can be seen that the tracks follow the same trend to within 
$\pm$0.02 up to about $r-i = 1.2$ (corresponding to M3 spectral type), after 
which there is a clear fanning out.  This shows that the obtained 
$r - H\alpha$ excesses should fall within the target photometric precision 
range of the survey for all except mid- to late-M stars.

The sensitivity of the M stars to variations in the narrowband 
transmission profile is a point of note, while not actually a surprise.  
It arises from the great breadth of the feature in M-star spectra created
by the absorbing TiO bands displaced to either side of the narrow H$\alpha$
bandpass, and the fact that the resulting inter-band flux maximum falls at 
wavelengths shortward of H$\alpha$.  As these molecular bands 
strengthen with increasingly late spectral type, the $r - H\alpha$ apparent 
excess grows along with the sensitivity to the exact placement of the 
bandpass.  Viewed in these terms, the $(r-H\alpha, r-i)$ colours of 
unreddened mid- to late-M dwarfs provide an empirical gauge of filter 
bandpass uniformity and/or typical CWL.  To minimise the bandpass 
sensitivity and hence spread seen at late M, the CWL would need to be 
lowered to around 6530~\AA\ or less.  In practice, later M dwarfs are 
sufficiently faint that they normally appear in $(r-H\alpha, r-i)$ diagrams 
as a relatively sparse distribution of scarcely reddened objects -- falling 
within a thinly-populated, continuation of the unreddened main
sequence, redward of $(r-i) = 1.0$, rising from $r - H\alpha \sim 0.5$
up to $\sim 0.8$, (see figs~\ref{fig:vphas-iphas}, or
\ref{fig:Wd2}).  Reddened M dwarfs are usually just too faint to
be detected.  

Selection of mid-to-late M dwarfs is therefore straightforward, but 
quantitative interpretation of $(r - H\alpha)$ should be presumed
more uncertain than at earlier spectral types.  Similar effects will be 
seen in the M-giant spur located at lower $(r - H\alpha)$ in the 
$(r-H\alpha, r-i)$ diagram (see fig~\ref{fig:hari-tracks}).  However, as 
red giants will be picked up by 
VPHAS+ at large distances through significant reddening, a precautionary 
check on the impact of non-zero extinction on this fanning in colour has 
been made: tracks of the type compared in fig~\ref{fig:track-comparison} 
were recalculated for $A_V = 6$ and no noticeable additional effect was 
found (see also fig~\ref{fig:vphas-iphas}).

\subsection{Simulation of the impact of source radial velocity on
  in-band emission line fluxes}
\label{sec:Ha-rv}

\begin{figure}
\centering{\includegraphics[angle=0,scale=0.45,trim=0 120 0 140]{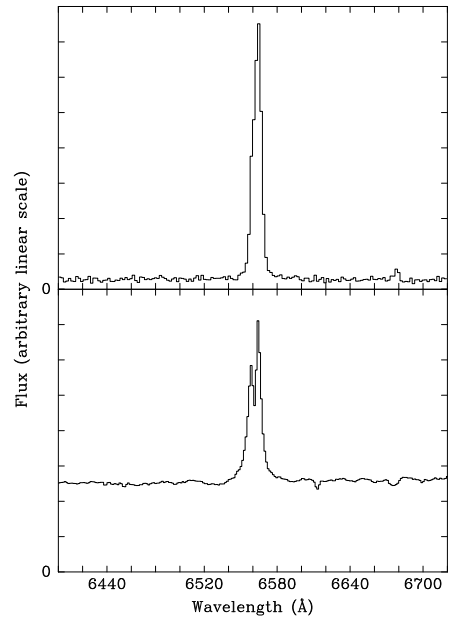}}
\caption {{\em Top panel:} an example of a very bright, simple H$\alpha$
emission profile (taken from Corradi et al 2010). The emission equivalent width
is 220~\AA , and the FWHM of the observed profile is close to 390 km~s$^{-1}$.
The mean radial velocity of the line is $+$35 km~s$^{-1}$.  The difference
between a pure continuum magnitude and that including the line is
somewhat in excess of 1.2. {\em Lower panel:} the contrast of the line
relative to continuum is much less here (EW $\sim$ 20 \AA ), and the 
FWHM is somewhat wider at 570 km~s$^{-1}$ (a classical Be star, taken
from Raddi et al, 2013).  The mean radial velocity is $-$50
km~s$^{-1}$.  The
continuum-only magnitude is fainter by about 0.2 only, here.} 
\label{fig:emission-spectra}
\end{figure}

Simulations have also been performed to consider how the filter
captures emitted $H\alpha$ flux, as a function of location within the field
of view and source radial velocity.  An ideal filter, centred on the mean 
rest wavelength of the imaged $H\alpha$ emission and placed in a high f-ratio 
optical path, would be insensitive to radial velocity shifts up to 
a limit proportional to the FWHM of the bandpass.  The desired capabilities 
of the VPHAS+ H$\alpha$ filter are separation between H$\alpha$ emission line 
objects and the main stellar locus -- and, better still, a regular mapping of 
measured $r - H\alpha$ excess onto emission equivalent width (cf  Drew et al 
2005, figure 6).  The two representative spectra used to investigate how these  
capabilities are affected by changing source radial velocity are shown in 
fig~\ref{fig:emission-spectra}.   

Both spectra were blueshifted to
$-$500 km~s$^{-1}$ , and then shifted redward in steps of 100 km~s$^{-1}$ at 
a time, up to $+$500 km~s$^{-1}$ (altogether a displacement of 22 \AA ) -- 
calculating at each step the integral of the spectrum folded through the 
filter transmission profile.  The resultant in-band fluxes were converted 
to magnitudes, and then shifted by the amount required to match the 
integrated transmission to the overall mean for the filter (again
mimicking the function of the pipeline illumination correction).  In
real use, we would expect the majority of emission line objects to
present with FWHM no greater than either of these examples
(interacting binaries and WR stars do present with much broader
emission, however).  

The radial velocity range explored was chosen with the following 
considerations in mind:-  Emission line stars in the thin disk will
commonly have radial velocities falling within the range -100 to +100
km/s.  In the Bulge larger radial velocities may be encountered:
excursions to $\pm$200 km~s$^{-1}$ are observed in CO (Dame, Hartmann
\& Thaddeus 2001) within $\sim 20^{\circ}$ of longitude of the
Galactic Centre, and for a minority of inner-Galaxy 
planetary nebulae, radial velocities have been obtained that extend
the range almost to $\pm$300 km~s$^{-1}$ (see Durand, Acker \& Zilstra 
1998, and Beaulieu et al 2000).

\begin{figure}
\centering{\includegraphics[angle=0,scale=0.5,trim= 0 160 0 180]{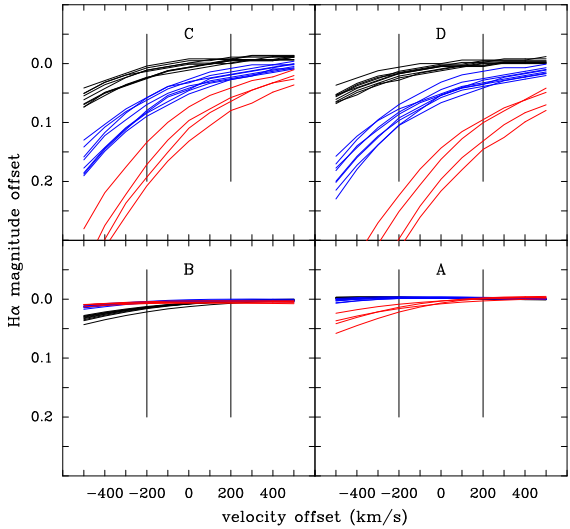}}
\caption {Results of simulation of the in-band flux as a function of source
radial velocity for the EW = 220~\AA\ emission spectrum (upper
panel in fig~\ref{fig:emission-spectra}).  The fluxes are expressed as
magnitude offsets relative to the peak simulated in-band flux. 
The panels representing the segments are arranged as they are imposed on the 
plane of the sky, i.e. A covers the SW quadrant, D covers the NW when the 
filter is stored in OmegaCAM's magazine B.  The curves are colour-coded
according to measurement position as in fig~\ref{fig:segment-map}.  
Galactic sources will usually fall well within the marked velocity range, 
$-$200 to $+$200 km~s$^{-1}$.  The more problematic red curves, 
representing the response of the segment corners, account for $\sim$8\% of 
each segment's unvignetted area only.}
\label{fig:high-contrast}
\end{figure}

\begin{figure}
\centering{\includegraphics[angle=0,scale=0.5,trim= 0 160 0 180]{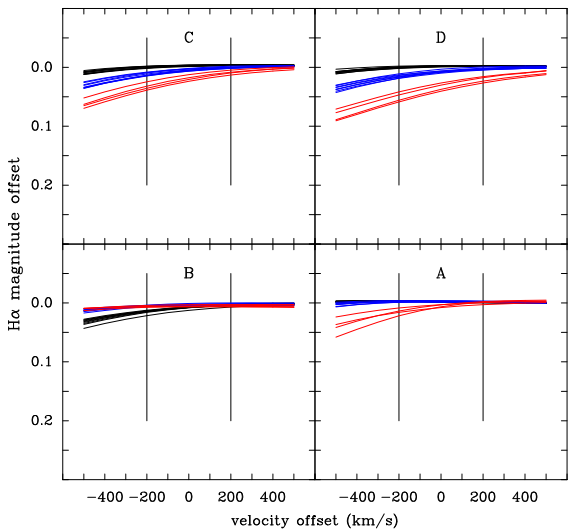}}
\caption {Results for an emission line net equivalent width of 
20 Angstroms (spectrum shown in lower panel in 
fig~\ref{fig:emission-spectra}).  Otherwise as 
fig~\ref{fig:high-contrast}.}
\label{fig:low-contrast}
\end{figure}

\begin{figure}
\centering{\includegraphics[angle=0,scale=0.5,trim= 0 240 0 130]{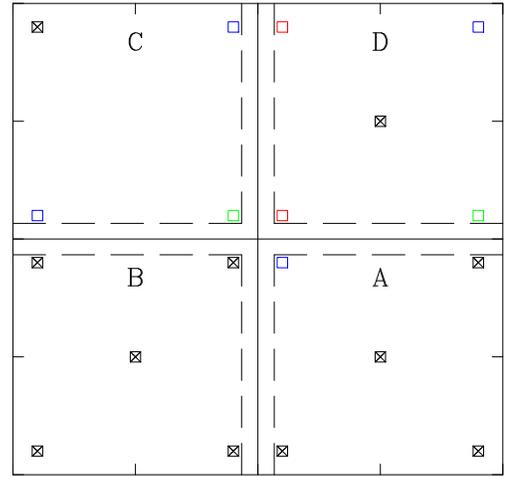}}
\caption {The layout of the H$\alpha$ filter, showing the 
positions of the PN-test measurements, colour coded according to 
flux relative to the corners of segment B. Those shown as black crossed
boxes have a scaled flux $\geq 0.95$.  Those in blue are a little lower 
with a scaled flux in the range 0.90 -- 0.94.  The two positions marked in 
green have fluxes scaling to 0.88 and 0.89, while the two in red scale
to 0.84 and 0.77 (both in segment D).  There is no measurement for the centre 
of segment C because of a telescope pointing imprecision.  The other centre
measurements were obtained $\sim 2$ arcmins away from the true 
centres plotted in order to avoid CCD gaps.  The dashed lines mark the
limits of the T-bar vignetting.}
\label{fig:PNtest}
\end{figure}

The results of this exercise are plotted in
figs~\ref{fig:high-contrast} and \ref{fig:low-contrast}.
Fig~\ref{fig:high-contrast} shows that segments A
and B come very much closer to independence of radial velocity than
segments C and D, in terms of measured $H\alpha$ excess.  In all
segments, reasonable fidelity (a flat, or nearly flat response) is
achieved around segment centre -- although in B, uniquely, the corners 
happen to perform a little better than the centre.  Clearly segment D, 
where the transmission is centred on longer wavelengths than in the
other segments, could 
yield measurements in its corners (perhaps 8\% of its unvignetted area) 
of $H\alpha$ magnitude, or of $r - H\alpha$, that underestimate the 
flux of similarly high equivalent-width H$\alpha$ emission by up to 
$\sim$ 0.3 (out of a true excess, expressed in magnitudes of $\sim
1.2$).  Segment C performs similarly, but the potential flux drop
associated with its corners is less pronounced.  
 
We can compare the expectations created by these simulations with the 
results of an on-sky experiment in which the planetary nebula (PN) ESO 178-5 
(or PNG 327.1 -02.2) has been exposed in H$\alpha$ at a series of positions in 
the image plane, placing it well into every corner and also close to
the centre of each of the four filter segments (see
fig~\ref{fig:PNtest}).  The observed integrated counts variation might
be predicted to be somewhat stronger than in
fig~\ref{fig:high-contrast} given that the 
in-band continuum flux from this PN will be relatively even weaker.  
But this will be offset by the additional flux due, in particular, to 
[NII]~$\lambda$6584.  The PN chosen for this test was picked both
because it is well-calibrated (Dopita \& Hua 1997), and because its 
LSR radial velocity is quite large and negative ($v_{LSR} =
-88.7$ km~s$^{-1}$: on 19th April 2013 when it was observed, this will
have shifted to $-105$ km~s$^{-1}$ at the telescope).  It also happens
to possess [NII]~$\lambda$6584 emission that is scarcely less bright
than H$\alpha$ (the former has 98\% the flux of the latter: Dopita \& 
Hua also provide a spectrum of this nebula).

Background-subtracted aperture photometry of the PN and a moderately 
bright star nearby, serving as a continuum reference, was carried out on 
the reduced images.  These measurements reveal a pattern of behaviour that 
essentially tracks the results shown in figure~\ref{fig:high-contrast}: the
continuum reference itself shows a total count variation of $\pm 5$\% 
across all pointings, while the PN counts, after scaling to the
reference, range from $+5$\%, down to $-23$\% relative to the values 
for the corners of segment B.  In the extreme case that all of the 
H$\alpha$ and [NII]~$\lambda$6548 emission had been shifted out of the 
bandpass, the maximum drop for this PN would be $-57$\% (the remaining
43\% being attributable to [NII]~$\lambda$6584).  The pattern across
the filter of the results emerging from this trial is shown in 
fig~\ref{fig:PNtest}.  

The radial-velocity dependence of the transmitted flux {\em may}
accordingly become an issue for objects with very strong H$\alpha$ 
emission where the aim is accurate flux determination, unless
attention is paid to where the object falls in the image plane.  
Qualitatively the issue is less critical: regardless of where the 
object is located, the changes in transmission are not so large that 
there will be frequent failures to distinguish strong H$\alpha$ emitters --
i.e. they will still appear above the main stellar locus in 
the $(r - H\alpha, r - i)$ diagram.   In the example simulated, the 
outer reaches of segment D would bring $(r - H\alpha)$ down to 0.9 -- 1.0,
a level that nevertheless remains clear of the domain that might
be occupied by unreddened, non-emission very late-type M dwarfs 
($(r - H\alpha) \sim 0.8)$, cf fig~\ref{fig:hari-tracks}).  As further
context, we note that nearly continuum-free emission line objects,
such as PNe and HII regions, present with $r - H\alpha \sim 3$.  

Where the line emission itself contributes only a minority of the
measured narrowband $H\alpha$ flux the trends seen are much more
subdued (fig~\ref{fig:low-contrast}).  For the example shown, only
20 percent of the total in-band flux is attributable to the net line 
emission, rather than most of it as in fig~\ref{fig:high-contrast}.  
Again, the corners of segments C and D perform least well, in 
under-representing the emission flux by up to $\sim$0.05 magnitudes at 
the most negative likely Galactic Plane radial velocities.  Otherwise, 
the performance is predicted to be within the anticipated 0.02--0.03 
error budget of the survey.

\subsection{Implications of the H$\alpha$ filter properties for VPHAS+ 
and its exploitation}
\label{sec:ha-final}

In summary, for most purposes the H$\alpha$ filter performs as
required, and has very good throughput.   
For the great majority of stars making up the main stellar locus, 
there will be the desired fidelity of $(r - H\alpha)$ colour, and
the great majority of emission line objects will be detected with the 
same facility as they are by IPHAS.  

There are two caveats to note.  First, in
Section~\ref{sec:MSHa-sim} it was shown that variations in central
wavelength across the filter segments will lead to thickening
of the loci traced by mid-to-late M stars.  These same variations, 
of what is a relatively red H$\alpha$ passband, also introduce the 
potential for under-determination of H$\alpha$ fluxes for 
objects/nebulosity in parts of the image plane for sources with
significantly negative radial velocities (Section~\ref{sec:Ha-rv}).  
This becomes most serious for emission-line sources falling near the 
vignetted corners of segments D and C, where a $20-30$ percent 
under-counting in H$\alpha$ may occur for radial velocities  
approaching $-200$~km s$^{-1}$.  As is always the case for narrowband 
H$\alpha$ filters, the common presence of significant [NII] 
$\lambda\lambda$6548, 6584 emission bracketing H$\alpha$ in planetary 
nebulae or HII regions complicates the expected signature.  However, 
it can be guaranteed in all but rare, exotic circumstances that the 
stronger $\lambda$6584 component of the [NII] doublet falls well
within the bandpass.

The outstanding practical consequence of the filter's transmission
characteristics for the survey strategy are that, for quantitative 
reliability, measures of $(r - H\alpha)$ obtained using segments A and
B, and the central zones of C and D (out to $\sim 13$ arcmin) are to 
be favoured.  This appreciation is half of the reason for the adoption
of offsets of several arcminutes between the 3 successive pointings
made in this filter (fig~\ref{fig:offset}) for each field -- our 
strategy ensures that objects captured in segment corners in the first 
pointing are fall close to segment centres in the third. 

The rest of the motive is to mitigate the cross-shaped
vignetting due to the blackened T-bars holding the segments in place 
(fig~\ref{fig:hafilter}).  Each arm of the cross casts a shadow 
{\em entirely} contained within a strip 4 arcmin wide.  By choosing to
offset at least this much in both RA and Dec between the 3 exposures obtained 
per field, we raise the probability of at least one high-confidence 
$r - H\alpha$ colour measurement per detected source in the final
catalogue to very nearly 100 percent, and the probability of two 
to over 95 percent.   

Finally, we remark that the combination of large offsets and three
pointings has the consequence that the fraction of sky within the
survey footprint missed altogether, due to dead areas between the CCDs
and vignetting, is under 0.3 percent.

\section{Simulation of VPHAS$+$ stellar colours}
\label{sec:simul}

   The five bandpasses of the survey provide the basis for the construction
of a range of magnitude-colour and colour-colour diagrams.  To take
full advantage of them, knowledge is needed of the behaviours that
can be expected of the colours of normal stars.

We have simulated colours for solar-metallicity main sequence and giant stars 
using the same method as employed by Sale et al. (2009). We adopt the 
definition of these two sequences in $(T_{eff}, \log g)$ space given by 
Straizys \& 
Kuriliene (1981). Then for each spectral type along these sequences, 
solar-metallicity model spectra were drawn from the Munari et
al. (2005) library.  At a binning of 
1~\AA\, the spectra in this library are well enough sampled to permit the 
calculation of narrow-band $H\alpha$ relative magnitudes with confidence, 
alongside the analogous broadband quantities.  More detail on 
the broadband filter transmission profiles, shown in Fig.~\ref{fig:filters}, 
and on the CCD response curve is provided on the ESO 
website\footnote{http://www.eso.org/sci/facilities/paranal/instruments/omegacam/doc}.  To ensure compliance with the 
Vega-based zero magnitude scale, we have defined the synthetic colour 
arising from a flux distribution $F_{\lambda}$ as follows:
\begin{displaymath}
  m_1 - m_2 = 
  -2.5 \log \left[\frac{\int T_{1} \lambda F_{\lambda} d\lambda}{\int T_{1}
          \lambda F_{\lambda ,V} d\lambda}\right] + 2.5 \log \left[
          \frac{\int T_{2}\lambda F_{\lambda} d\lambda}{\int T_{2}
          \lambda F_{\lambda ,V} d\lambda}\right]
\label{eqn_colour}
\end{displaymath} 
where $T_{1}$ and $T_{2}$ are the numerical transmission profiles
for filters $1$ and $2$, after multiplying them through by the 
atmospheric transmission (Patat et al 2011) and mean OmegaCAM CCD response 
curves.  The SED adopted for Vega, $F_{\lambda ,V}$, is that due to Kurucz
(http://kurucz.harvard.edu/stars.html).  Where needed for comparison, we 
have also computed colours based on the Pickles (1998, hereafter P98) 
spectrophometric 
stellar library (the approach adopted by Drew et al 2005 for IPHAS).  To
maintain precision, the numerical quadrature resamples
the more smoothly varying transmission data onto the sampling interval 
of the stellar SED. 

The $(r - H\alpha)$ excess is evaluated in exactly the same way as the 
broadband colours.  Since Vega is an A0V star, its SED at H$\alpha$ 
incorporates a strong absorption line feature that reduces the in-band 
flux below the pure continuum value.  Unlike the broadbands, the $H\alpha$ 
narrowband has not yet been standardised and so there is not a formally
recognised flux scale.  However, we can specify here that the
integrated in-band energy flux for Vega, on adopting the mean profile 
for the VST filter, is $1.84\times10^{-7}$ ergs~cm$^{-2}$~s$^{-1}$ (at
the top of the Earth's atmosphere).  To assure zero colour relative to 
the optical broad bands, this flux is required to correspond to
$m_{H\alpha} \simeq 0.03$.  The reduction 
in zeropoint (zpt) that the computed in-band flux implies relative
to the flux captured by the much broader $r$ band -- based on folding
Vega's SED with lab measurements of the filter throughputs corrected 
for atmosphere and detector quantum efficiency -- is 3.01. Current 
practice in VPHAS+ photometric calibration is accordingly to adopt 
zpt(NB-659) = zpt(r) - 3.01 magnitudes as the default calibration for 
the narrowband: in section~\ref{sec:phot-val} where a direct
comparison is made with SDSS spectroscopy, this offset is found to
be satisfactory.  When applied, it assures that data obtained in 
photometric, or stable, conditions, yield zero $r - H\alpha$ colour 
for A0 stars.


\begin{figure}
\centering{\includegraphics[angle=0,scale=0.5,trim= 0 220 0 220]{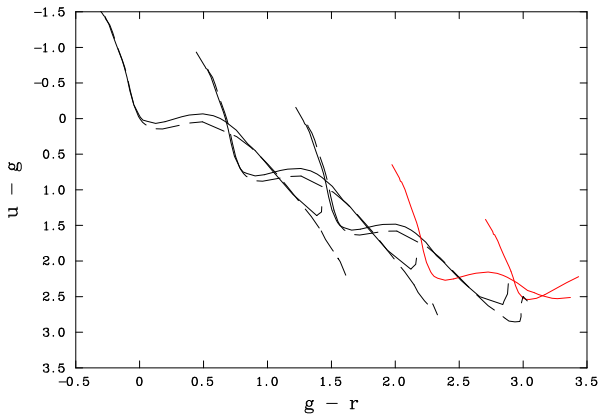}}
\caption{The expected positions of main sequence and 
giant stars in the $(u - g, g - r)$ plane.  For the main sequence, 
tracks are shown for the monochromatic reddenings $A_0 =$ 0, 2, 4, 6
and 8 (working from left to right).  The red leak in the $u$ filter
starts to lower $u - g$, noticeably from $A_0 = 6$ (tracks drawn in red). The 
giant-star tracks, drawn as dashed lines for $A_0 = 0$, 2 and 4 only, are 
very similar to their main-squence counterparts except at the latest 
types.} 
\label{fig:ugr-tracks}
\end{figure}

\begin{figure}
\centering{\includegraphics[angle=0,scale=0.5,trim= 0 220 0 220]{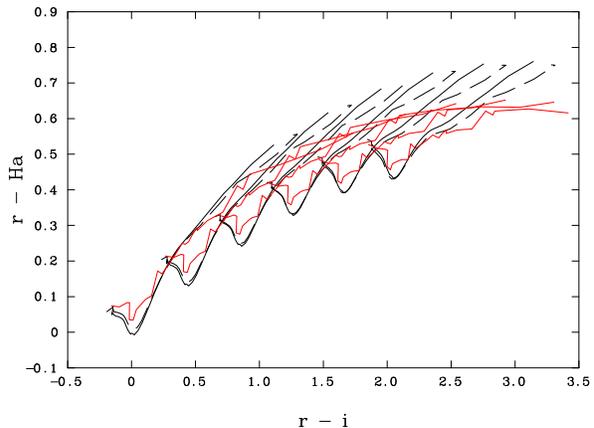}}
\caption{The expected position of main sequence and giant stars in the
  $(r - H\alpha, r - i)$ plane.  Tracks are shown for the
  monochromatic reddenings $A_0 =$ 0, 2, 4,
  6, 8 and 10 (from left to right).  Solid lines represent the 
  main sequence tracks, while dashed 
  lines are used for the giant tracks.  The lines in red are giant-star 
  tracks derived from P98 spectrophotometry.}
\label{fig:hari-tracks}
\end{figure}

Both main-sequence and giant-star colours have been calculated for a 
range of reddenings and optical-IR extinction 
laws as formulated by Fitzpatrick \& Massa (2007).  The
unreddened colours for the mean Galactic law ($R = 3.1$) are laid out 
here in table~\ref{tab:colours0}.   The Appendix provides additional 
tables that specify the colours of main sequence stars at selected reddenings 
and for two further representative reddening laws ($R = 2.5$ and 3.8).  
These can be used to construct intrinsic-colour-specific reddening
lines.  For the large range in extinction sampled along many Galactic
Plane sightlines, these reddening trends are slightly curved (see the 
examples shown in e.g. Sale et al 2009).  In this paper we use $A_0$,
the monochromatic reddening at 5500~\AA\ to parameterise the amount of
reddening, rather than the band-averaged measure, $A_V$.  In most 
circumstances these quantities are almost identical.

\begin{table*}
\caption{VST/OmegaCAM synthetic colours for unreddened main-sequence dwarfs 
and giants.}
\begin{tabular}{cccccccccccc}
\hline
Sp.type & \multicolumn{4}{c}{main sequence (V)} & \ & \multicolumn{6}{c}{giants (III)} \\
   &    &    &    & \ &   &    \multicolumn{4}{c}{(model)} &
\multicolumn{2}{c}{(P98 library spectra)} \\
        & $u - g$ & $g - r$ & $r - i$ & $r - H\alpha$ & \ & $u - g$ & $g - r$ & $r - i$ & $r - H\alpha$ & $r - i$ & $r - H\alpha$ \\
\hline 
  O6 & -1.494 & -0.313 & -0.145 &  0.071 & \ &       &        &        &       &        & \\
  O8 & -1.463 & -0.299 & -0.152 &  0.055 & \ &       &        &        &       & -0.158 & 0.074 \\
  O9 & -1.426 & -0.271 & -0.142 &  0.064 & \ &-1.426 & -0.271 & -0.142 & 0.064 &        & \\
  B0 & -1.404 & -0.267 & -0.143 &  0.058 & \ &-1.404 & -0.267 & -0.143 & 0.058 &        & \\
  B1 & -1.296 & -0.236 & -0.130 &  0.052 & \ &-1.316 & -0.234 & -0.130 & 0.057 & -0.095 & 0.071 \\
  B2 & -1.181 & -0.214 & -0.117 &  0.049 & \ &-1.209 & -0.211 & -0.116 & 0.056 &        & \\
  B3 & -1.025 & -0.182 & -0.098 &  0.048 & \ &-1.046 & -0.182 & -0.098 & 0.054 & -0.035 & 0.083 \\
  B5 & -0.799 & -0.133 & -0.071 &  0.043 & \ &-0.814 & -0.134 & -0.072 & 0.050 & -0.016 & 0.083 \\
  B6 & -0.699 & -0.116 & -0.062 &  0.040 & \ &-0.714 & -0.116 & -0.062 & 0.046 &        & \\
  B7 & -0.550 & -0.094 & -0.051 &  0.033 & \ &-0.568 & -0.095 & -0.051 & 0.041 &        & \\
  B8 & -0.361 & -0.071 & -0.039 &  0.022 & \ &-0.383 & -0.072 & -0.039 & 0.032 &        & \\
  B9 & -0.168 & -0.040 & -0.023 &  0.009 & \ &-0.186 & -0.044 & -0.024 & 0.021 & -0.018 & 0.035 \\
  A0 & -0.024 &  0.000 & -0.003 & -0.002 & \ &-0.030 & -0.009 & -0.006 & 0.011 & 0.012 & 0.034 \\
  A1 &  0.007 &  0.015 &  0.004  & -0.004 & \ & 0.007 & 0.004 &  0.000 & 0.008 &       & \\
  A2 &  0.039 &  0.038 &  0.014  & -0.005 & \ & 0.051 & 0.022 & 0.009  & 0.008 &       & \\
  A3 &  0.064 &  0.062 &  0.025  & -0.005 & \ & 0.085 & 0.045  & 0.019  & 0.007 & 0.037 & 0.063 \\
  A5 &  0.096 &  0.130 &  0.056  &  0.008 & \ & 0.143 & 0.107  & 0.048  & 0.015 & 0.096 & 0.087 \\
  A7 &  0.073 &  0.206 &  0.089  &  0.030 & \ & 0.145 & 0.179  & 0.078  & 0.032 & 0.115 & 0.094 \\
  F0 &  0.003 & 0.336  &  0.153  &  0.086 & \ & 0.091 & 0.317  & 0.144  & 0.084 & 0.156 & 0.102 \\
  F2 & -0.021 & 0.396  &  0.182  &  0.111 & \ & 0.064 & 0.380  & 0.174  & 0.109 & 0.204 & 0.172 \\
  F5 & -0.039 & 0.505  &  0.230  &  0.150 & \ & 0.046 & 0.491  & 0.224  & 0.148 & 0.238 & 0.164 \\
  F8 & -0.013 & 0.587  &  0.263  &  0.174 & \ &       &        &        &       &       & \\
  G0 &  0.012 & 0.628  &  0.278  &  0.185 & \ &       &        &        &       & 0.333 & 0.213 \\
  G2 &  0.011 & 0.628  &  0.279  &  0.185 & \ & 0.253 & 0.759  & 0.329  & 0.215 &       & \\
  G5 &  0.162 & 0.756  &  0.327  &  0.217 & \ & 0.405 & 0.870  & 0.368  & 0.235 & 0.396 & 0.250 \\
  G8 &  0.355 & 0.845  &  0.358  &  0.233 & \ & 0.531 & 0.944  & 0.395  & 0.247 & 0.419 & 0.247 \\
  K0 &  0.523 & 0.938  &  0.396  &  0.248 & \ & 0.640 & 1.002  & 0.417  & 0.256 & 0.446 & 0.254 \\
  K1 &  0.551 & 0.954  &  0.403  &  0.251 & \ & 0.803 & 1.085  & 0.451  & 0.269 & 0.468 & 0.269 \\
  K2 &  0.629 & 0.993  &  0.419  &  0.258 & \ & 0.963 & 1.159  & 0.483  & 0.281 & 0.508 & 0.293 \\
  K3 &  0.779 & 1.062  &  0.447  &  0.269 & \ & 1.227 & 1.276  & 0.536  & 0.299 & 0.514 & 0.286 \\
  K4 &  0.871 & 1.108  &  0.468  &  0.278 & \ & 1.374 & 1.342  & 0.585  & 0.320 & 0.592 & 0.313 \\
  K5 &  1.083 & 1.210  &  0.522  &  0.300 & \ & 1.578 & 1.420  & 0.630  & 0.338 & 0.714 & 0.337 \\
  K7 &  1.387 & 1.402  &  0.724  &  0.387 & \ &       &        &        &       &       & \\
  M0 &  1.372 & 1.411  &  0.789  &  0.411 & \ & 1.697 & 1.454  & 0.686  & 0.360 & 0.827 & 0.411 \\
  M1 &  1.335 & 1.439  &  0.934  &  0.467 & \ & 1.838 & 1.506  & 0.769  & 0.395 & 0.872 & 0.401 \\
  M2 &  1.262 & 1.442  &  1.112  &  0.522 & \ & 1.938 & 1.546  & 0.825  & 0.420 & 0.920 & 0.443 \\
  M3 &  1.236 & 1.447  &  1.179  &  0.545 & \ & 1.980 & 1.556  & 0.933  & 0.444 & 1.165 & 0.471 \\
  M4 &  1.248 & 1.457  &  1.168  &  0.543 & \ & 1.959 & 1.551  & 1.136  & 0.500 & 1.472 & 0.512 \\
  M5 &        &        &         &        & \ & 2.009 & 1.569  & 1.296  & 0.556 & 1.739 & 0.560 \\
  M6 &        &        &         &        & \ & 2.199 & 1.612  & 1.267  & 0.554 &       & \\
\hline
\end{tabular}
\label{tab:colours0}
\end{table*}

Based on the data from these tables, the main sequence and giant
tracks are as shown in figures~\ref{fig:ugr-tracks} and
\ref{fig:hari-tracks}.  These identify where the main stellar loci will
fall.  It is important to note that the OmegaCAM $u$ filter, like all
filters constructed for this challenging band, exhibits a low-level 
red leak.  In this instance, lab measurements indicate transmission at
levels between $10^{-5}$ and $10^{-4}$ within limited windows around 
$\sim$9000~\AA .  This is enough to begin to noticeably, and
erroneously, brighten the $u$ magnitudes of normal stars reddened to 
$g - r > 3$.  Because of this, and because the measurement of very
low level leakage is itself subject to proportionately higher
uncertainty, we do not plot or tabulate $u - g$ data beyond $g - r
= 3$ limit.  Very few detected sources are so extreme.  In practice, 
VPHAS+ $u - g$ is faithful for extinctions up to $A_0 \sim 6$, but 
gradually thereafter it transforms into a colour that behaves
crudely as $-(g - z)$.

The $(r - H\alpha, r - i)$ colour-colour diagram is not subject
to such effects, and therefore remains sound across a wider spread in visual 
extinctions.  Synthetic tracks are presented in figure~\ref{fig:hari-tracks} 
for $A_0 = 0$, 2, 4, 6, 8 and 10.  The main sequence tracks shown are  
similar to those appropriate to IPHAS (cf figure 6 of Sale et al 2009).  But a
problem emerges when it comes to the simulation of red giant colours.
Purely theoretical simulation predicts late-K and M giant colours closely 
resembling those of dwarfs, whereas simulation using P98 library 
spectrophotometry indicates a distinctive flattening of the M-giant
track, peeling away from the steadily rising main sequence track.  
Figure~\ref{fig:hari-tracks} points out this contrast.  Inspection of 
table~\ref{tab:colours0} reveals this is a problem linked mainly to 
simulation of the $i$ spectral range, which renders $(r - i)$
progressively larger from late K into the M giant range when the
library spectra are used in place of model atmospheres.

\begin{figure}
\centering{\includegraphics[angle=0,scale=0.6,trim= 60 160 0 220]{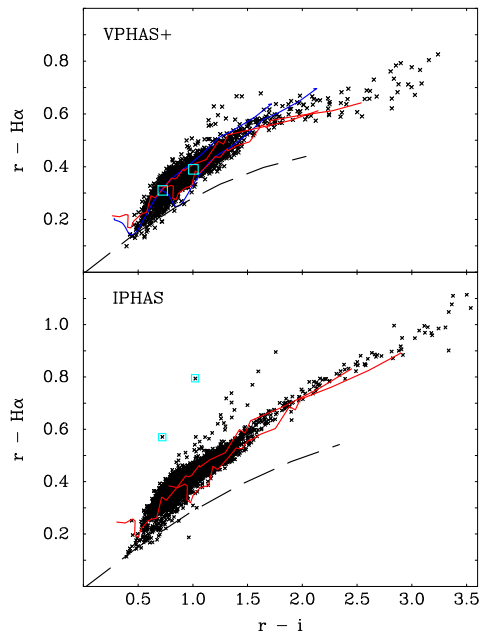}}
\caption{Equatorial VPHAS+ data (upper panel) and IPHAS data (lower
  panel) compared to show the different appearance of M-giant
  $(r - H\alpha, r - i)$ colours. The photometry
  is extracted from a $\sim$0.2 sq.deg region of sky, centred on
  $\ell = 35.95^{\circ}$, $b = -3.13^{\circ}$.  The magnitude range is
  limited to $13 < r < 18$ in both cases.  Telescope-appropriate
  giant tracks, computed from the P98 spectrophotometric
  library, for $A_0 =$ 2 and 4 are superimposed in red in
  both panels.  In the upper panel, giant tracks computed from model
  atmospheres for the same reddenings are shown in blue. The black
  dashed line in each panel is the synthetic early-A reddening line,
  from $A_0 = 0$ to $A_0 = 10$.  
  The stars appearing in the IPHAS catalogue as candidate emission
  line stars are enclosed in cyan boxes in both panels.}
\label{fig:vphas-iphas}
\end{figure}

Evidence that M giants are better reproduced by synthetic photometry
based on flux-calibrated spectra is provided by figure~\ref{fig:vphas-iphas}.  
This figure also compares new VPHAS+ data with their crossmatches in
the IPHAS survey within a $\sim$0.2 sq.deg equatorial field ($\ell = 
35.95^{\circ}$, $b = -3.13^{\circ}$), and shows selected synthetic tracks
superimposed.  The photometry from the two surveys of the most densely 
populated part of the main stellar locus to $(r - i) \simeq 1.5$
substantially overlap, but not perfectly -- the response functions
describing the three bandpasses involved in fig~\ref{fig:vphas-iphas}
undoubtedly differ in detail between the two telescopes.  On
cross-correlating either $(r - i)$ or $(r - H\alpha)$ between the two
surveys, it becomes clear that the IPHAS colour has the somewhat larger
dynamic range.  This is the reason for the slightly more stretched
appearance of both the main locus and the early-A reddening line in 
the IPHAS diagram relative to that for VPHAS+. 

At $(r - i) > 1.5$ in fig~\ref{fig:vphas-iphas}, it can be
seen that the M-giant spurs look very different.  First, the VPHAS+ 
M giants fall into a nearly flat distribution lying at lower 
$(r - H\alpha)$, compared to the more steeply rising higher IPHAS
M-giant sequence.  However, as long as the data are interpreted 
with reference to telescope-appropriate synthetic photometry,
the two datasets will lead to the same inference.  In the example
shown in fig~\ref{fig:vphas-iphas}, the comparisons with suitable
synthetic giant tracks indicate that the maximal extinction in the
field can be no more than $A_0 \simeq 4$.  The extinction measures due to
Marshall et al (2006), based on 2MASS red-giant photometry, indicate a 
maximum Galactic extinction of $A_K \sim 0.3$ for this pointing.  For
a typical Galactic $R = 3.1$ reddening law this scales up to $A_0 \sim
3.3$ (roughly -- see Fitzpatrick \& Massa 2009).  If model-atmosphere 
giant tracks are referred to instead, the M giants would have to be 
read as demanding visual extinctions ranging from $\sim$4 upwards.  

Fig~\ref{fig:vphas-iphas} also demonstrates the broadening in 
$(r - H\alpha)$ of the VPHAS+ M-giant sequence that was foretold in 
Section~\ref{sec:ha}.  The IPHAS counterpart is evidently much
sharper, as it rises to higher $r - H\alpha$ with increasing 
$r - i$. The main practical impact of this difference is that IPHAS M-giant
photometry is the better starting point for picking apart chemistry 
differences (Wright et al 2008).  But it is as true of the VPHAS+ 
$(r - H\alpha, r - i)$ diagram as it is of its IPHAS equivalent -- 
that M giants at $r - i > 1.5$ sit below and apart from M dwarfs. 

Fortuitously, fig~\ref{fig:vphas-iphas} identifies an advantage of
the generally good seeing available at the VST.  There are two 
candidate emission line objects apparent in the IPHAS selection
(enclosed in cyan boxes, in fig~\ref{fig:vphas-iphas}), that drop 
back into the main stellar locus in the VPHAS+ data.  Inspection of
the images shows that both stars are in close doubles of similar 
brightness, of under 2 arcsec separation.  Because they are a little 
better resolved in VPHAS+ ($\sim$0.8~arcsec seeing), than in IPHAS 
($\sim$0.9~arcsec seeing), the pipeline makes a better job of the 
assigning magnitudes in the different bands to the blend components.
This example nicely illustrates the most common reason for bogus 
candidate emission line stars in either VPHAS+ or IPHAS - improperly
disentangled blends.  Candidate emission line stars should always be 
checked for this kind of problem before spectroscopic follow-up.
Otherwise, experience with IPHAS gives confidence that the selection of
emission line objects via VPHAS+ will be highly efficient (see e.g.
Vink et al 2008, Raddi et al 2013). 

Finally, it is worth noting that the bright limit of the survey at
12--13th magnitude effectively excludes any unreddened stars of earlier
spectral type than $\sim$G0.  Before more luminous stars of spectral
type F and earlier can enter the survey sensitivity range, they need
to be at distances in excess of 1 kpc, typically, where low
extinction becomes increasingly improbable.  This constraint bestows a 
significant selection benefit in that only unreddened or
lightly-reddened subluminous objects, with intrinsically blue colours
are left standing clear near the blue end of the main stellar locus in
commonly-constructed photometric diagrams.  In this domain VPHAS+ has 
important selection work to do.

\section{Photometry validation: a comparison of SDSS and VST data}
\label{sec:phot-val}


Before the start of survey field acquisition, we obtained observations
in all survey filters of two pointings that fall within the SDSS 
photometric and spectroscopic coverage 
(Abazajian et al 2009).  These were centred on RA 20:47:53.7 Dec -06:04:14.5
(J2000) and RA 21:04:25.94 Dec +00:59:15.8 (J2000) -- fields that
happen to include a number of white dwarfs and cataclysmic variables
(not discussed further here).  The main aim of the data was to
verify VST photometry both by comparison to SDSS photometry and to 
synthetic photometry derived from SDSS spectra.  
The VST observations were obtained on 21/09/2011, during
clear weather at a time of generally sub-arcsecond seeing.  The
exposure times differ only a little from those now in general survey 
use: the $g$ exposures were 30 sec, rather than 40, and $i$ was
exposed for 20 sec rather than 25 - the other times were as 
given in Section~\ref{sec:survey-def}.  

The photometry on the sources in these fields have been 
pipeline-extracted and calibrated in the standard way, and have
been cross-matched to their SDSS counterparts.  The number of cross 
matched stars used in this exercise ranges from $\sim 2500$ ($u$) up to 
$\sim 10000$ ($r$).  For a star to be included, it must be:
unvignetted; to have a star-like point spread function; to lie within
0.5 arcsec of its SDSS counterpart, and to fall within the magnitude 
range $16 > r_{VST} > 19$.  The SDSS selection constraints were set to 
exclude blended and saturated sources, and sources close to detector
edges. In addition, it was required of every source that, in both
surveys, the formal error on the magnitude measurement is less than 
0.03. 

\begin{figure}
\centering{\includegraphics[angle=0,scale=0.5,trim= 40 170 -30 220]{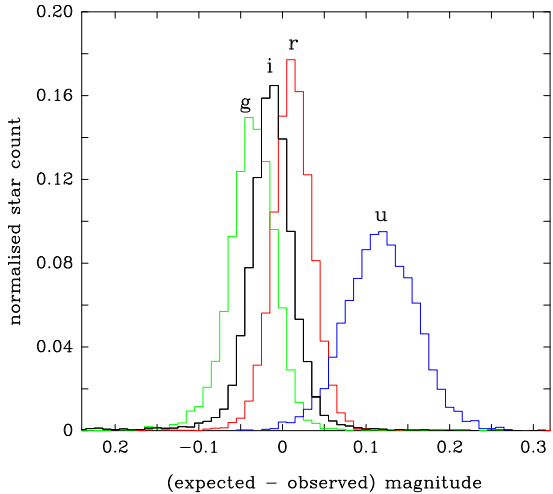}}
\caption{The distributions, by band pass, of the measured magnitude 
  differences between VST test data and SDSS, after correction of the 
  latter magnitudes from the AB to the Vega zero-magnitude scale.
  These were obtained from two VST pointings, away from the Galactic
  Plane, that overlap the SDSS footprint.  The biggest differences are
  found in the $u$ band where the VST data are fainter than the
  corrected SDSS values by $\sim 0.12$, in the median.} 
\label{fig:sloan-hist}
\end{figure}

\begin{figure}
\centering{\includegraphics[angle=0,scale=0.45,trim= 40 150 10 200]{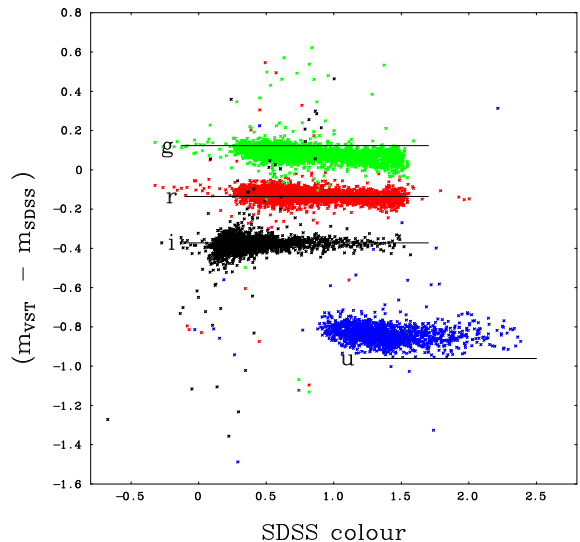}}
\caption{Measured magnitude shifts between VST and SDSS cross-matched
objects as a function of SDSS colour, for the second of the two fields
observed (only).  The data in blue are $\Delta u$
versus $(u - g)_{SDSS}$; the data in green and red are, respectively, 
$\Delta g$ and $\Delta r$ versus $(g - r)_{SDSS}$; the data in black are 
$\Delta i$ versus $(r - i)_{SDSS}$.  The horizontal lines show where
the four loci would be expected to lie in the case that the SDSS and 
VST broadband filters were identical, and the calibrations perfect.  
The data fall into loci that are not far displaced from these
horizontal lines and are almost as flat: this indicates that the 
colour-dependent terms that would be needed in equations to transform 
between the SDSS and VST systems are small.} 
\label{fig:sloan-colour}
\end{figure}


In fig~\ref{fig:sloan-hist} we plot the histograms of the 
magnitude differences between the two surveys, according to pass 
band -- pooling the data from both pointings.  If the starting
assumption is that the VST broadband filters are identical to the SDSS
set, the predicted magnitude for each star in each filter is 
the measured SDSS magnitude, less the offset between the AB and 
Vega scales (essentially the numerical difference between the
magnitude of Vega in the AB system and its value of 0.02 to 0.03, 
according to the alternative Vega-based convention -- see table 8 
in Fukugita et al 1996).  In the $g$, $r$ and $i$ bands, the 
predicted and observed magnitudes are well-enough aligned, and the 
interquartile spread is consistent with the way the data were selected 
for random errors less than 0.03. However in $u$ there is a 
discrepancy that exceeds expected error: the median difference 
deviates by 0.12 magnitudes and the width of the distribution is 
twice that arising in the comparisons of the other bands. 

The fuller picture is presented in figure~\ref{fig:sloan-colour} which 
shows the broadband magnitude differences as a function of the
relevant SDSS colour for the second of our two fields (only). In all 4
pass bands, including $u$, the colour dependence can be seen to be very
weak in that the loci traced out by the plotted stars are -- to a first
approximation -- flat.  The discrepancy seen in $u$ is revealed as mainly 
a zero-point shift, combined with scatter that exceeds the formal
errors.  In highly reddened Galactic Plane fields, the stellar colour
effects may become more pronounced as extinction modifies the
effective sampling of the passbands.

\begin{table}
\caption{Mean magnitude offsets between VST photometry and cross-matching 
  synthetic photometry derived from the SDSS database of spectra.  The 
  synthetic magnitude scale adopts magnitudes for Vega itself of 0.026 
 (see Bohlin \& Gilliland 2004)}
\begin{tabular}{lcc}
\hline 
Offset & Field 1 (117 stars) & Field 2 (50 stars) \\
\hline
$u_{VST} - u_{syn}$ & 0.07$\pm$0.39 & 0.11$\pm$0.17 \\  
$g_{VST} - g_{syn}$ & 0.06$\pm$0.09 & 0.06$\pm$0.04 \\ 
$r_{VST} - r_{syn}$ & 0.01$\pm$0.06 & 0.03$\pm$0.03 \\
$i_{VST} - i_{syn}$ & -0.06$\pm$0.08 & -0.05$\pm$0.03 \\ 
\hline
\end{tabular}
\label{tab:sloan-comp2}
\end{table}

As a separate exercise, we have used SDSS spectra to synthesise 
magnitudes and colours for stars with cross-matching VST photometry.  
The spectral type range present within this much smaller sample runs from 
B-type through to early M-type (M1).  At wavelengths below 3800~\AA\ 
falling within the $u$ band, it was necessary to extrapolate the
spectra using appropriately chosen P98 library data.  The result of
this comparison is agreement between the VST and synthesised
magnitudes at the $\sim$5 percent level (table~\ref{tab:sloan-comp2}), 
with the $u$ band as the outlier exhibiting much more pronounced
scatter as well as somewhat higher offset.  This pattern
echoes the behaviour apparent in the VST-SDSS purely photometric 
comparison of fig~\ref{fig:sloan-hist}, using a much larger sample.  
The difficulty is not confined to VST $u$ however, in that SDSS $u$ 
photometry fares scarcely any better relative to synthesis from the 
spectra (for the two fields, offset and scatter are -0.09 $\pm$ 0.37, 
and -0.04 $\pm$ 0.16).   As more blue survey data are accumulated, it 
may become clear that the $u$ zeropoint will benefit from being tied
to that of $g$ for those fields observed in the best conditions, as is 
presently done for $H\alpha$ with respect to $r$.  This option is not 
yet enacted.  For the time being, it must be acknowledged that 
pipeline $u$ calibration is more approximate than those of the other bands.

We have also used the reduced cross-match sample to look at how the 
VST photometric $r-H\alpha$ colour compares with its
counterpart synthesised from spectra -- looking, in particular, for any 
trends as a function of distance from field centre.  No such trend 
is apparent, thereby meeting the expectation that the narrowband
fluxes of normal stars, to early-M spectral type, would not be 
affected by the pattern of bandpass shifts discussed earlier in 
section~\ref{sec:ha} (cf. fig~\ref{fig:track-comparison}).  
However we do find that in order to make this detailed comparison, 
systematic offsets had to be removed from the VST photometry first.  
These were 0.075 in $(r - i)$, in the sense that the VST
colours were too red by this amount, and a 0.02 reduction in $(r -
H\alpha)$.  The $(r - i)$ offset is consistent with the 
broadband magnitude offsets listed in Table~\ref{tab:sloan-comp2} and 
hence is as expected.  The $(r - H\alpha)$ adjustment is small 
enough (i.e. within the fit error) that it supports the zeropoint shift
of 3.01 magnitudes between the $r$ and $H\alpha$ bands that was
identified in section~\ref{sec:simul}).  Once these colour offsets 
are applied to the VST data, the rms scatter of the photometric 
$(r - H\alpha)$ colour relative to its synthetic counterpart is 0.04 
for objects brighter than $r = 19$.

\begin{figure*}
\centering{\includegraphics[angle=0,scale=0.5]{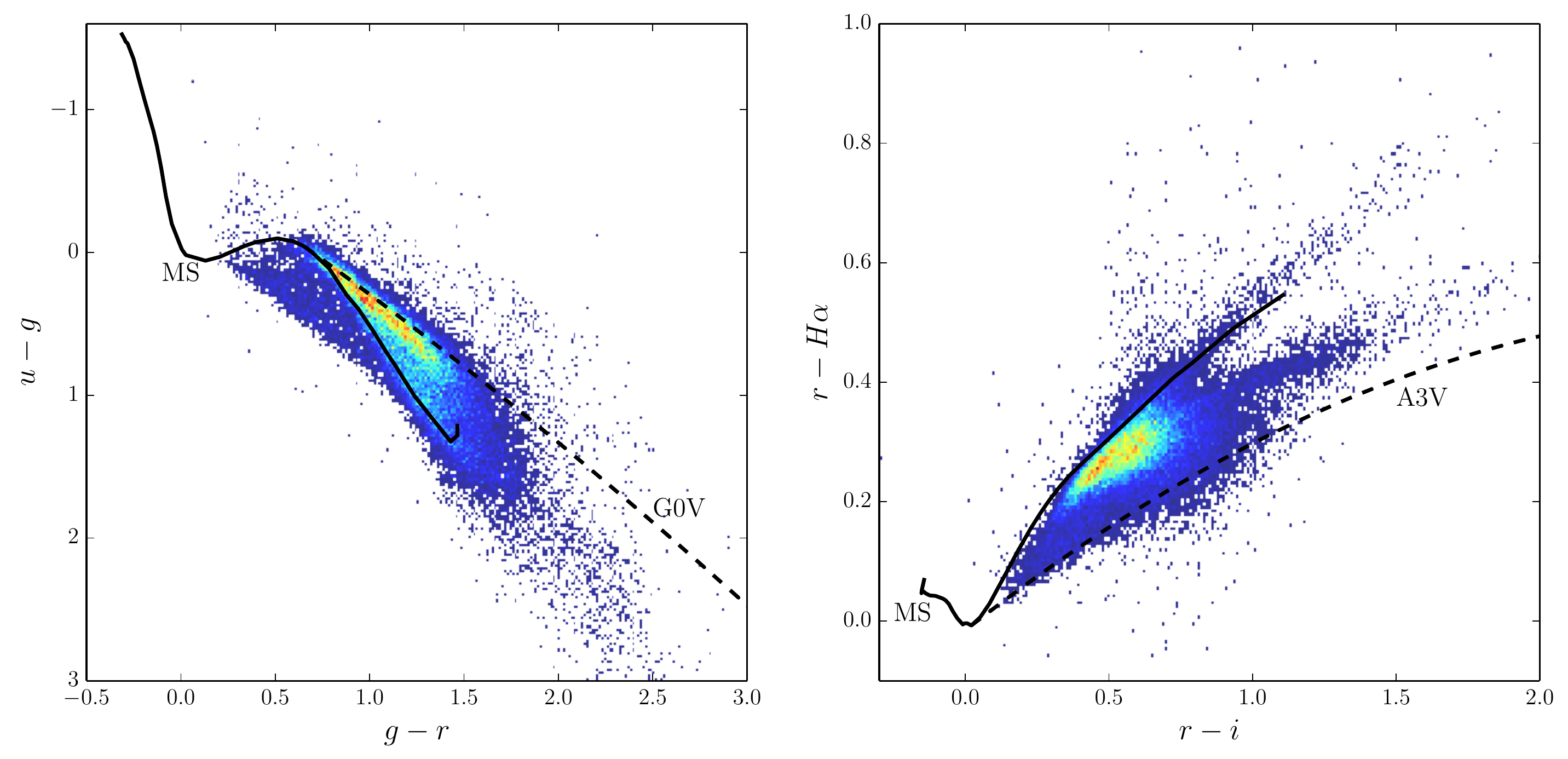}}
\caption{Left, the $(u-g, g-r)$ and right, the $(r-H\alpha, r-i)$ 
  photometric diagrams pertaining to VPHAS+ survey field 1679.  
  Both diagrams are plotted as
  two-dimensional stellar-density histograms, rainbow colour-coded 
  such that high source densities (80-90 per bin) are red and the
  lowest densities (one per bin) are dark blue. The binning is
  0.017$\times$0.025 in the left panel, 0.013$\times$0.008 in the
  right.  The
  synthetic unreddened main sequence is drawn in, in black, in both
  panels. The G0V and A3V reddening lines obtained for 
  $R = 3.1$, drawn as black dashed lines, are included in respectively 
  the $(u-g, g-r)$ and $(r-H\alpha, r-i)$ diagrams as useful aids to 
  interpretation.} 
\label{fig:Wd2}
\end{figure*}


\section{An example of point-source photometry derived from a VPHAS+
  field}
\label{sec:example} 

The extracted point-source photometry from the VST square-degree 
field is one of the two main data products from VPHAS+ -- the other
being the images themselves, considered below in
section~\ref{sec:imaging}.  We present an example of the two essential 
colour-colour diagrams in fig~\ref{fig:Wd2}, in which band-merged stellar 
photometry for field 1679 is compared with the primary diagnostic 
synthetic tracks presented in section~\ref{sec:simul}.  This field 
includes the sky area from which data were taken to construct 
fig~\ref{fig:errors} illustrating typical errors.  The massive open
cluster, Westerlund 2, is located in the NE of these pointings, and
the field as a whole 
includes moderate levels of diffuse, complex H{\sc ii} emission.   
The data presented are drawn from a sky area centred on RA 10 24 49 
Dec -57 58 00 (J2000) that spans 1.3 sq.deg -- the total footprint 
occupied by the two offset positions.

Only objects with stellar point-spread functions in $g$, $r$ and $i$, 
brighter than $g = 20$ are included in fig~\ref{fig:Wd2}.  Where two
sets of magnitudes are
available, the mean values have been computed and used.  A further
requirement imposed is that the random error in all bands may not 
exceed 0.1.  The same $\sim$37000 objects are included in both
diagrams.  In order to obtain the diagrams shown, the pipeline
photometric calibration was checked and refined as follows: we
\begin{itemize}
\item cross-matched brighter stars to APASS $g$, $r$ and $i$ photometry
\item computed the median magnitude offset (applying no colour corrections
-- it was shown in fig~\ref{fig:sloan-colour} these are modest) 
\item corrected all $g$, $r$ and $i$ for these offsets; 
\item corrected $u$ by determining the vertical shift needed in the 
$(u-g, g-r)$ diagram to align the main stellar locus with the
  unreddened main sequence and the G0V reddening line
\item corrected the $H\alpha$ zeropoint and hence all $H\alpha$
  magnitudes according to the requirement that $zpt(H\alpha) = zpt(r)
  - 3.01$.
\end{itemize}
This resulted in the following broadband corrections:- $\Delta i =
-0.004$, $\Delta r = -0.032$ (red filter set), $\Delta r = -0.033$
(blue filter set), $\Delta g = 0.069$ and $\Delta u = -0.31$. 
As expected, the correction that had to be applied to the
$u$ photometry was, by far, the largest.

The main stellar locus can be seen to be tightly concentrated in both
the blue and the red diagrams, and to favour lightly reddened G and K
stars.  The superimposed synthetic reddening lines (G0V in the $(u-g,
g-r)$ diagram, A3V in $(r-H\alpha, r-i)$) have been drawn adopting the
$R = 3.1$ reddening law widely regarded as the Galactic norm.  
The blue diagram provides examples of three distinct typical
populations falling outside the main stellar locus.  Below it, at
$(u-g) > 1.5$ and $(g-r) > 1.5$ (roughly) the plotted objects will
mainly be M giants.  Above the main stellar locus toward the red end,
in the ranges $0 < (u-g) < 0.5$ and $1.5 < (g-r) < 2.0$ lie the OB
stars in and around Westerlund 2.  Finally, the modest scatter of blue 
objects lying above the G0V line roughly in the $0 \leq (g-r) \leq
1$ range will include intrinsically-blue lightly-reddened subluminous 
objects.

It is interesting to note in the red diagram that there is some evidence
that early-A stars making up the lower edge of the main stellar locus
would better follow a different law, with $R \sim 3.8$ (see the tables
in the Appendix).  Indeed a reddening law of this type has been
inferred for the OB stars in Westerlund~2 by Vargas Alvarez et al 
(2013).  Most of the thin scatter of points below the main stellar 
locus, and some of the scatter above, in this same diagram will be the 
product of inaccurate background subtraction in H$\alpha$.  But many
of the objects lying above the main stellar locus will indeed be
emission line objects, and some of the stars below will be white
dwarfs.  As expected, the red spurs of M dwarfs and
M giants are broader features than their IPHAS counterparts (cf 
fig~\ref{fig:vphas-iphas} and associated remarks).

For more discussion of these colour-colour diagrams, the reader is
referred to Groot et al (2009, UVEX) and Drew et al (2005, IPHAS). 

\begin{figure*}
\centering{\includegraphics[angle=0,scale=0.7]{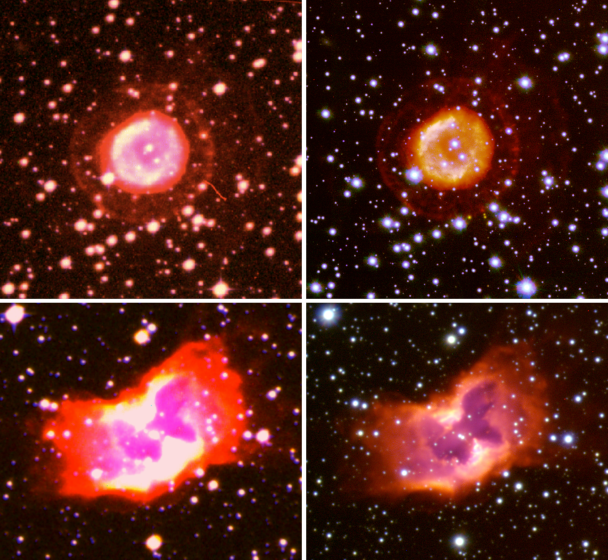}}
\caption{Two planetary nebulae, NGC 2438 (top) and NGC 2899 (bottom), as 
  they appear in the SHS and VPHAS+ surveys. The SHS images are shown in the 
  left-hand panels, with the VPHAS+ images to the right. The bands
  used to form them are: NGC 2438 -- SHS R/G/B = $H\alpha$/SR/SSS~Bj, 
  VPHAS+ R/G/B = $H\alpha$/$r$/$i$, NGC 2899 -- SHS R/G/B =
  $H\alpha$/SR/SSS~Bj, VPHAS+ R/G/B = $H\alpha$/$r$/$g$.  The cut-out image 
  dimensions are 300$\times$300 arcsec$^2$ for NGC 2438, and 200$\times$180 
  arcsec$^2$ for NGC 2899.
} 
\label{fig:pne}
\end{figure*} 

\section{Nebular astrophysics with VPHAS+ images}
\label{sec:imaging}


Just over a decade ago the SuperCOSMOS H$\alpha$ Survey (SHS, Parker et al
2005) had only just completed.  This was the last survey using 
photographic emulsions that the UK Schmidt Telescope undertook.  The
3-hour narrowband H$\alpha$ filter exposures reach a very similar
limiting surface brightness to the 2 minute exposures VPHAS+ is built around.
Hence, the differences in capability are not about sensitivity, as
this is roughly the same in the two surveys.  Instead it is about the
great improvement in dynamic range on switching to digital detectors,
the good seeing of the VST's Paranal site, and the added broad bands.  

SHS, with its enormous 5-degree diameter field, has been comprehensively 
trawled for southern planetary nebulae (the MASH catalogue, Parker et al 
2006, Miszalski et al 2008).  
The remaining discovery space for resolved nebulae is expected to be
at low surface 
brightnesses in locations of high stellar density, and in the compact domain
around and below the limits of the typical spatial resolution of 
SHS ($\sim 0.5$ to 3.0 arcsec).  Both these conditions will most often
be met in the Galactic Bulge, at a mean distance of $\sim$8 kpc.  
Data-taking in the Bulge and its maximally-dense star fields is planned to 
begin in mid 2014.   

Among planetary nebulae (PNe), small angular size is due either to
great distance or to youth -- the study of either compact category 
provides exciting possibilities.  
As well as the Bulge, the less-studied outer parts of the Galactic
Plane should be searched.  In this respect, IPHAS, 
with its direct view to the Galactic Anticentre is better positioned:
the ongoing study of the Anticentre PN population has revealed dozens
of new candidates (Viironen et al. 2009a), including the PN with the 
largest galactocentric distance to date (20.8 $\pm$ 3.8 kpc, Viironen et 
al 2011).  By following up such finds to measure chemical abundances,
crucial beacons are obtained for the study of the Galactic abundance
gradient and its much disputed flattening towards the largest
galactocentric radii.  VPHAS+ completed the access to the outer Plane 
over the longitude range $215^{\circ} < \ell < 270^{\circ}$.

Data from both IPHAS and VPHAS+ can make fundamental contributions to
the study of very young PNe -- particularly by helping to solve the 
two-decades-old puzzle of how PNe already emerge with the observed wide 
variety of morphologies (round, elliptical, bipolar, multipolar, 
point-symmetric, etc. -- see Sahai et al. 2011).  What does this
variety say about the properties of their AGB progenitors? Detailed 
studies of objects in the phases preceding the PN phase -- AGB and
post-AGB stars, proto-PNe, and transition or PN-nascent objects -- are 
underway (e.g. Sanchez Contreras \& Sahai 2012).  Superb imaging 
capabilities like those of the VST, accessed via VPHAS+, will support
this work. 

Indeed there is a serious paucity of very small PNe in the existing 
optical catalogues: there are no PNe with angular extent less than 3 
arcsec in the MASH catalogue (out of 903 objects; Parker et al 2006), 
and only 8 PNe in the catalogue by Tylenda et al (2003, 312 objects) 
in the size range 1.4 -- 3 arcsec. There is just one with a 
confidently-measured diameter below 1 arcsec in the larger Strasbourg 
Catalogue of PNe (1143 objects;  Acker et al. 1994), that happens to 
be a Bulge PN.  IPHAS has demonstrated that extremely young compact
PNe can be reached (Viironen et al, 2009b), while Sabin et al (in
prep.) have found some 20 new PNe with diameters of 1-3 arcsec in 
by-eye searchs of IPHAS image mosaics.  Even smaller, but brighter, 
nebulae around symbiotic stars of the dusty D subtype are emerging 
-- the record so far being IPHASJ193943.36+262933.1, a new D symbiotic star
with an $H\alpha$ extent of only 0.12 arcsec that has been confirmed via 
HST imaging and recently studied with the 10.4m GTC telescope 
(Rodriguez Flores et al. 2014, submitted to A\&A).

Apart from opening up new discoveries, a further benefit of good
seeing is the clearer view of nebular structure that it offers.  This
is nicely demonstrated in fig~\ref{fig:pne}.  SHS and VPHAS+
detect the main features of the planetary nebulae NGC 2438 and NGC
2899 to very similar depth -- for
example, the fainter outer halo is just detected in both versions of 
NGC 2438. But, evidently, the VPHAS+ images better resolve the fine
sculpting within both nebulae as a consequence of the seeing FWHM
being under a half that prevailing in SHS data.  The extended 
dynamic range of VPHAS+ helps in this respect, too, in that early
saturation also obliterates detail.  This advantage is especially clear in 
the images of NGC 2899, where the structure in the bright nebulous lobes is 
preserved in VPHAS+, but is entirely bleached out in SHS.  The more
the level of detail that can be picked out, the more certain and
subtle morphological classifications and interpretations can become.

\begin{figure}
\centering{\includegraphics[angle=0,scale=0.6,trim= 0 0 0 400]{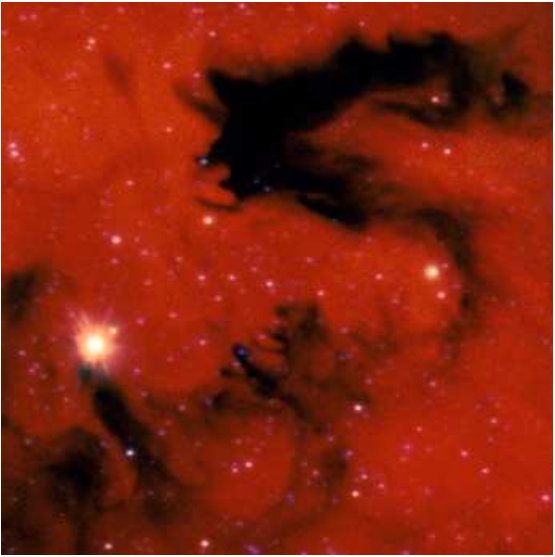}}
\caption{A cut-out at full resolution from M8, the Lagoon Nebula.
  This is an RGB image centred on RA 18 09 36 Dec -24 01 51 (J2000),
  and spanning 150$\times$150 arcsec$^{2}$.
  The filters are combined such that R/G/B = $H\alpha$/$i$/$r$.} 
\label{fig:hii-detail}
\end{figure}

The combination of good seeing and high dynamic range also makes the 
separation of fainter stars from background nebulosity much easier.  
This capability is critically important to the study of the
young massive clusters, still swathed in diffuse H{\sc ii} emission, 
where the analysis of stellar content is very much a focus of
continuing research. For example, Feigelson et al (2013) have offered
a critique of the nuisance created by spatially-complex
nebulosity. The obvious
answer to this and the problem of dust obscuration is to turn to
selection using NIR and X-ray data.  Nevertheless the availability of 
imaging data of the high quality seen in VPHAS+ data will make it possible
to extend SEDs for many more stars into the effective-temperature (and
reddening) sensitive optical domain. In addition, understanding the 
shaping of the interstellar medium in star-forming environments remains
an important part of the picture (see e.g. Wright et al 2012 on proplyd-like
structures in Cyg OB2). The detail that the VST is capable
of revealing both in obscuration and ionised hydrogen in star-forming 
regions can be quite exquisite.  Here, in fig~\ref{fig:hii-detail}, we 
illustrate this with an excerpt from VPHAS+ data on the Lagoon Nebula, 
showing the fine tracing of the shapes of dark globules and eroding dusty
structures that is achieved.

\begin{figure}
\centering{\includegraphics[angle=0,scale=0.7,trim=0 0 0 200]{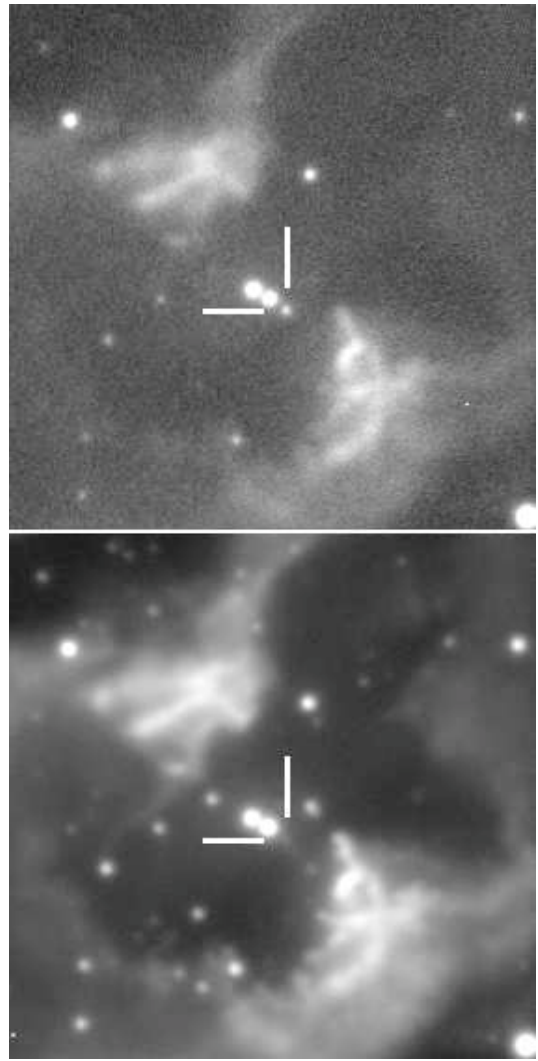}}
\caption{The central star of NGC 2899 reveals itself.  The top panel
  is a 1$\times$1 arcmin$^2$ thumbnail of the centre of NGC 2899 as
  imaged through the $u$
  filter, while the bottom is the corresponding $r$ thumbnail.  The white
  right-angled bars pick out the position of an extremely blue,
  relatively faint star that is clearly present in all $u$ (and $g$)
  exposures obtained, but is too faint for detection in $r$.} 
\label{fig:ngc2899-cs}
\end{figure}

In planetary and other evolved-star nebulae it is of course important
to identify the ionising object.  The
search for missing PN central stars is a quest that VPHAS+ can aid 
greatly through the provision of spatially well-resolved $u$ and $g$ 
data.  Indeed inspection of the data used to
contruct fig~\ref{fig:pne} has revealed the probable central star of 
NGC 2899 for the first time.  As shown in fig~\ref{fig:ngc2899-cs},
there is very evidently a third very-blue star just SW of the pair of
stars that have, in the past, been scrutinised as possible companions
to what is required to be an extremely hot ($T_{eff} > 250,000$ K), but
probably faint central star (L\'opez et al 1991).  This blue object was
detected on the night of 20th December 2012 at a provisional $u$ magnitude 
of 18.79 $\pm 0.02$. It fades through $g$ (19.36 $\pm 0.02$) to become 
undetected by the pipeline, and scarcely visible to eye inspection, in $r$.  
Its coordinates are RA 09 27 02.72 Dec -56 -06 22.9 (J2000), just 1.7~arcsec 
from the more southerly of the pair of brighter stars examined before by 
L\'opez et al (1991). Based on the $g$ magnitude and an inferred $V$ flux, we 
have determined the central star's effective temperature, via the 
well-established Zanstra method.  Using the reddening and integrated 
H$\alpha$ flux from L\'opez et al and Frew, Boji\v{c}i\'c \& Parker 
(2013) respectively, we estimate $T_{\rm z,H} = 215 \pm 16$~kK.  This is
cooler than the temperature given by L\'opez et al, based on
the 'crossover' method, but still extraordinarily hot for a central
star well down the white-dwarf cooling track.

It was one of the major science drivers 
for the merged VPHAS+ survey that $u$ data, supported by $g$, would
result in the detection of a broad range of intrinsically very blue
objects -- be they PN central stars, interacting binaries or massive
OB and Wolf-Rayet stars.  An extreme example like NGC 2899's central
star provides the useful lesson that selection via the $u-g, g-r$
colour-colour diagram would have failed to pick it out -- because of
the non-detection in $r$.  In a case like this, the $u,u-g$ 
colour-magnitude diagram has to be examined, in tandem with
the appropriate images.

\section{VPHAS+ photometry as a reference set for variability studies}
\label{sec:var}


As the northern survey, IPHAS, has progressed over the decade since
2003, there have been occasions on which it was possible to
use the growing database as a high-quality reference for 
checking transient reports -- particularly of novae.  The most spectacular
IPHAS example of this was the nova and variable, 
V458~Vul (Wesson et al 2008; Rodriguez-Gil et al 2010) where the
eruption occurred a few months after obtaining H$\alpha$ images revealing
a pre-existing ionised nebula around the star.  Indeed there have been
several instances in which photometry of the progenitor object has
been extracted from the IPHAS database and has been used to gain
insight into the prior presence or absence of line emission or to set
constraints on likely extinction (Steeghs et al 2007, Greimel et al
2012).   Such opportunities will certainly arise with VPHAS+ 
-- and be richer given the five filters offered.  

In the southern hemisphere novae will be more frequent, as will other 
transient events.  Furthermore
responses to alerts, or the need to demonstrate long-term flux
variations, can bring into use repeats of observations made necessary 
by initial quality-control failures. An example of
this is provided by Vink et al (2008) who used repeat IPHAS observations 
-- taken on account of poor observing conditions -- to discuss the LBV 
candidacy of G79.29$+$0.46.  
With the increased attention being given to the reporting and
exploitation of transient objects (including the forthcoming Gaia
alerts programme), this use of VPHAS+ will become more common.


\section{Summary and concluding remarks} 
\label{sec:last}

This paper has introduced and defined VPHAS$+$, the VST Photometric 
$H\alpha$ Survey of the Southern Galactic Plane and Bulge. The data
taking, the rationale behind it, the data processing and data quality
have all been described.  The properties and limitations of the 
survey's narrowband H$\alpha$ filter, NB-659, have been 
laid out and simulated in order to anticipate its performance.  
In addition we have provided tables of the expected photometric
colours of normal solar-metallicity stars to aid the 
interpretation of the survey's characteristic photometric diagrams -- 
most are to be found in the Appendix where the effect of changing
the adopted reddening law is illustrated.  The VPHAS+ H$\alpha$ filter
transmission is redder, wider and $\sim20$ \% higher-throughput than
its IPHAS counterpart -- a difference that feeds through to noticeably 
different $(r - H\alpha)$ colours for M stars. 

We have validated the photometry that is delivered by VST/OmegaCAM and 
subsequently pipelined at CASU, using test data taken of a field for
which SDSS photometry is available. We find the agreement is
satisfactory, with the $g$, $r$ and $i$ band calibrations differing by 
between 0.01 to 0.05 magnitudes.  However, for the time being, the
pipeline calibration should be regarded as provisional -- it will
undoubtedly improve.  Examples of the excellent imaging performance of 
the VST/OmegaCAM combination relative to previous surveys have been
provided, and we draw attention to the valuable archival role this
first digital survey can fulfill in supporting discoveries of transient 
sources. 

Exploitation of the survey is now beginning.  The detection of a compact 
ionized nebula around W26, the extreme M supergiant in Westerlund 1 has 
already been published (Wright et al 2014).  Applications have been made 
for follow-up spectroscopy that will test the quality of
selection of specialised object types that VPHAS+ photometry makes
possible. Progress is also being made via direct analyses of the photometry.  
For example, Mohr-Smith et al (in prep.) are conducting a search for OB
stars in the vicinity of the massive cluster Westerlund 2, and they are 
finding a close match between the properties of known cluster O stars as 
derived from VPHAS+ data and those inferred by Vargas Alvarez et al
(2013, see also Drew et al 2013).  
This and other early appraisals of the data indicate that VPHAS+ will be an 
excellent vehicle for automated searches for reddened early-type 
stars.  Kalari et al (in prep.) are employing both narrow-band 
H$\alpha$ and the broadbands to measure mass-accretion rates in 
pre-main-sequence stars: they are finding that H$\alpha$ mass-accretion
rates in T~Tauri stars compare favourably to rates determined from
the $u$ band in the case of the Lagoon Nebula, NGC~6530.  

As the calibration of the survey data improves, the measurement of accurate
integrated H$\alpha$ fluxes for many faint southern PNe and other extended
objects becomes possible, and will extend the work of Frew et al
(2013, 2014).  In due course these fluxes can be compared with
existing and also new radio continuum fluxes coming on stream (see
e.g. Norris et al 2011) in order to determine reliable extinction
values for many faint nebulae currently lacking data.  This technique 
has already been applied to the case of W26 in Westerlund 1 (Wright et
al 2014).

When it becomes possible to cross-match VVV and VPHAS$+$ data, it will open 
up the power of homogeneous photometric mapping of the central parts of the 
Galactic Plane in up to 10 photometric bands spanning both the optical
and the near-infrared. Beyond the VVV sky area, there is a synergy to be 
exploited in bringing VPHAS+ data together with those of the all-sky 2MASS 
survey (Skrutskie et al 2006) and with the UKIDSS Galactic Plane Survey 
(Lucas et al 2008), in those parts of the first and third Galactic quadrants 
the latter has covered. It is worth noting, however, that 2MASS alone is too 
shallow to link effectively with VPHAS+ for sightlines where the integrated 
visual extinction is less than $\sim 5$ magnitudes.  This does mean that the 
longitude range $230^{\circ} < \ell < 300^{\circ}$, in particular, is presently 
lacking sufficiently deep NIR photometry.  In the longer
term, many of the sources of interest that VPHAS+ finds will benefit
from accurate parallaxes and other data from ESA's Gaia mission --
given the similar sensitivity limits reached.  Conversely in the
meantime, VPHAS+ has already begun to assist ambitious wide-field 
spectroscopy programmes such as the Gaia-ESO Survey (Gilmore et al 2012) 
through the provision of the wide-field photometry needed for target 
selection and field setup.

By the end of 2013, 25\% of all observations making up the survey had
been obtained to the required quality, and in May 2013 a first release of
single-band catalogues was made to the ESO archive that contained
roughly 10\% of the eventual total (based on data 
obtained prior to 15 October 2012).  By design, the characteristics of VPHAS+
are similar to those of the IPHAS and UVEX Galactic plane survey
pair in the north.  In particular, the double-pass strategy is shared,
with the result that the majority of detected objects are picked up and 
measured twice, with no more than $\sim$0.1 percent of objects missed 
altogether.  This feature has informed the way in which the IPHAS DR2
catalogue (Barentsen et al, 2014) has been constructed -- and it is
intended that a first band-merged VPHAS+ catalogue, for public release, will 
be built along analogous lines during the second half of 2014.  This will
incorporate data from the first 3 seasons of VST observing, and give a 
complete photometric account of the Galactic mid-plane.  For ease of use,
for every detected source, the catalogue will provide a single
recommended set of magnitudes in up to 5 optical bands.

\section*{Acknowledgments}
This paper makes use of public survey data (programme 177.D-3023) 
obtained via queue observing at the
European Southern Observatory.  In respect of the H$\alpha$ filter, we 
would very much like to thank Bernard Muschielok for the benefit of his 
expertise and support in connection with its laboratory testing, and 
Jean-Louis Lizon for his steady hand in correcting some minor surface
defects.  The referee of this paper is thanked for constructive 
comments that improved its content.

This research made use of the AAVSO Photometric All-Sky Survey (APASS), 
funded by the Robert Martin Ayers Sciences Fund.  Many elements of
the data analysis contained in this work have been eased greatly 
by the TOPCAT package created and maintained by Mark Taylor (Taylor,
2005).  The pipeline reduction also makes significant use of data from 
the Two Micron All Sky Survey (2MASS), which is a joint project of the
University of Massachusetts and the Infrared Processing and Analysis
Center/California Institute of Technology, funded by NASA and the NSF.  

JED and GB acknowledge the support of a grant from the Science \& Technology 
Facilities Council of the UK (STFC, ref ST/J001335/1). The research leading 
to these results has also benefitted from funding from the European 
Research Council under the European Union's Seventh Framework
Programme (FP/2007-2013) / ERC Grant Agreement n. 320964 (WDTracer).  
BTG was also supported in part by the UK STFC (ST/I001719/1).  
RLMC and AMR acknowledge funding from the Spanish 
AYA2007-66804 and AYA2012-35330 grants. HJF and 
MM-S both acknowledge STFC postgraduate studentships. NJW is in
receipt of a Royal Astronomical Society Fellowship.  RW acknowledges funding 
from the Marie Curie Actions of the European Commission (FP7-COFUND).

\appendix

\section{Synthetic colour reddening tables} 

Synthetic colours for main sequence stars, computed as described in 
Section~\ref{sec:simul}, are tabulated in full in an online supplement for 
three representative reddening laws ($R_V = 2.5$, 3.1 and 3.8) and a 
range of reddenings ($A_0 = 0$, 2, 4, 6, 8, 10).  The form of the 
reddening laws used is due to Fitzpatrick \& Massa (2007).  
As an example of the tables available we include excerpts from the 
second and fifth tables that respectively provide $R_V = 3.1$
blue-filter and red-filter colours for B,A main-sequence stars.


Two further tables of synthetic colours are included in the supplement 
for K-M giants that have been computed using P98 library spectra.
Data are provided for the $R = 3.1$ mean Galactic law only, for the 
limited purposes of (a) giving an impression of how these luminous 
red objects may 
contaminate $(u-g, g-r)$ diagrams at redder $(g-r)$ through $u$ red
leak (b) enabling comparisons with the M-giant spur commonly seen in 
$(r-H\alpha, r-i)$ colour-colour diagrams. 

\newpage

\begin{table*}
\caption{VST/OmegaCAM synthetic colours for B,A main-sequence stars in the
  $(u - g),(g - r)$ plane reddened with an $R_V=3.1$ extinction
  law. (Full table online.)}
\begin{tabular}{c c c c c c c c c c c c} 
\hline
\\*[-5pt]
Spectral &\multicolumn{2}{c}{$A_0=0$} & \multicolumn{2}{c}{$A_0=2$}&\multicolumn{2}{c}{$A_0=4$}&\multicolumn{2}{c}{$A_0=6$}&\multicolumn{2}{c}{$A_0=8$}\\
Type & $(u - g)$&$(g - r)$&$(u - g)$&$(g - r)$&$(u - g)$&$(g - r)$&$(u - g)$&$(g - r)$&$(u - g)$&$(g - r)$\\
\\*[-5pt]
\hline\\
$B0V$&$-1.433$&$-0.271$&$-0.692$&$0.529$&$0.087$&$1.301$&$0.891$&$2.050$&$1.632$&$2.777$\\
$B1V$&$-1.324$&$-0.240$&$-0.584$&$0.558$&$0.195$&$1.329$&$0.995$&$2.076$&$1.719$&$2.802$\\
$B2V$&$-1.209$&$-0.218$&$-0.470$&$0.579$&$0.307$&$1.350$&$1.104$&$2.096$&$1.808$&$2.821$\\
$B3V$&$-1.053$&$-0.186$&$-0.315$&$0.610$&$0.460$&$1.379$&$1.250$&$2.125$&$1.923$&$2.849$\\
$B5V$&$-0.828$&$-0.139$&$-0.092$&$0.655$&$0.680$&$1.423$&$1.460$&$2.166$&$2.080$&$2.890$\\
$B6V$&$-0.728$&$-0.121$&$0.007$&$0.672$&$0.776$&$1.439$&$1.550$&$2.182$&$2.144$&$2.905$\\
$B7V$&$-0.580$&$-0.100$&$0.152$&$0.692$&$0.918$&$1.458$&$1.682$&$2.200$&$2.234$&$2.922$\\
$B8V$&$-0.388$&$-0.076$&$0.340$&$0.714$&$1.101$&$1.478$&$1.850$&$2.219$&$2.344$&$2.940$\\
$B9V$&$-0.198$&$-0.046$&$0.528$&$0.742$&$1.285$&$1.504$&$2.019$&$2.244$&$2.445$&$2.964$\\
$A0V$&$-0.053$&$-0.005$&$0.675$&$0.780$&$1.431$&$1.540$&$2.153$&$2.277$&$2.514$&$2.995$\\
$A1V$&$-0.019$&$0.005$&$0.709$&$0.790$&$1.464$&$1.550$&$2.181$&$2.287$&$2.525$&$3.005$\\
$A2V$&$0.021$&$0.025$&$0.749$&$0.809$&$1.505$&$1.568$&$2.217$&$2.304$&$2.538$&$3.021$\\
$A3V$&$0.038$&$0.059$&$0.771$&$0.840$&$1.531$&$1.597$&$2.241$&$2.332$&$2.541$&$3.048$\\
$A5V$&$0.067$&$0.125$&$0.805$&$0.904$&$1.567$&$1.658$&$2.269$&$2.390$&$2.523$&$3.105$\\
$A7V$&$0.044$&$0.199$&$0.788$&$0.975$&$1.554$&$1.726$&$2.252$&$2.456$&$2.474$&$3.169$\\
\\
\hline
\end{tabular}
\end{table*}

\begin{table*}
\caption{VST/OmegaCAM synthetic colours for B,A main-sequence stars in the
  $(r - i),(r - H\alpha)$ plane reddened with an $R_V=3.1$ extinction
  law. (Full table online.)}
\begin{tabular}{c c c c c c c c c c c c c c} 
\hline
\\*[-5pt]
Spectral &\multicolumn{2}{c}{$A_0=0$} & \multicolumn{2}{c}{$A_0=2$}&\multicolumn{2}{c}{$A_0=4$}&\multicolumn{2}{c}{$A_0=6$}&\multicolumn{2}{c}{$A_0=8$}&\multicolumn{2}{c}{$A_0=10$}\\
Type & $(r - i)$&$(r - H\alpha)$&$(r - i)$&$(r - H\alpha)$&$(r - i)$&$(r - H\alpha)$&$(r - i)$&$(r - H\alpha)$&$(r - i)$&$(r - H\alpha)$&$(r - i)$&$(r - H\alpha)$\\
\\*[-5pt]
\hline\\
$B0V$&$-0.150$&$0.054$&$0.278$&$0.198$&$0.694$&$0.316$&$1.100$&$0.409$&$1.496$&$0.478$&$1.884$&$0.526$\\
$B1V$&$-0.136$&$0.048$&$0.291$&$0.192$&$0.708$&$0.310$&$1.114$&$0.403$&$1.510$&$0.472$&$1.898$&$0.519$\\
$B2V$&$-0.123$&$0.045$&$0.304$&$0.188$&$0.721$&$0.306$&$1.126$&$0.398$&$1.523$&$0.466$&$1.911$&$0.513$\\
$B3V$&$-0.104$&$0.044$&$0.323$&$0.186$&$0.740$&$0.303$&$1.145$&$0.394$&$1.541$&$0.462$&$1.929$&$0.508$\\
$B5V$&$-0.077$&$0.039$&$0.349$&$0.180$&$0.765$&$0.295$&$1.170$&$0.386$&$1.566$&$0.452$&$1.954$&$0.497$\\
$B6V$&$-0.068$&$0.036$&$0.358$&$0.177$&$0.774$&$0.291$&$1.179$&$0.381$&$1.575$&$0.448$&$1.963$&$0.492$\\
$B7V$&$-0.057$&$0.029$&$0.369$&$0.170$&$0.785$&$0.284$&$1.190$&$0.374$&$1.586$&$0.440$&$1.973$&$0.484$\\
$B8V$&$-0.045$&$0.018$&$0.382$&$0.158$&$0.797$&$0.272$&$1.202$&$0.362$&$1.598$&$0.427$&$1.985$&$0.471$\\
$B9V$&$-0.028$&$0.006$&$0.398$&$0.145$&$0.813$&$0.259$&$1.218$&$0.348$&$1.614$&$0.413$&$2.001$&$0.456$\\
$A0V$&$-0.009$&$-0.005$&$0.418$&$0.133$&$0.833$&$0.246$&$1.238$&$0.334$&$1.633$&$0.399$&$2.020$&$0.441$\\
$A1V$&$-0.003$&$-0.003$&$0.423$&$0.135$&$0.838$&$0.248$&$1.243$&$0.335$&$1.638$&$0.399$&$2.025$&$0.442$\\
$A2V$&$0.006$&$-0.004$&$0.432$&$0.134$&$0.847$&$0.247$&$1.251$&$0.334$&$1.646$&$0.397$&$2.033$&$0.439$\\
$A3V$&$0.021$&$-0.008$&$0.446$&$0.130$&$0.861$&$0.241$&$1.265$&$0.328$&$1.660$&$0.391$&$2.047$&$0.432$\\
$A5V$&$0.051$&$0.005$&$0.476$&$0.141$&$0.890$&$0.250$&$1.293$&$0.335$&$1.687$&$0.396$&$2.073$&$0.436$\\
$A7V$&$0.083$&$0.027$&$0.507$&$0.160$&$0.920$&$0.268$&$1.322$&$0.350$&$1.716$&$0.410$&$2.101$&$0.448$\\
\\
\hline
\end{tabular}
\end{table*}

\end{document}